# Nucleosynthetic osmium isotope anomalies in acid leachates of the Murchison meteorite

L. Reisberg[1], N. Dauphas[2], A. Luguet[3,4], D.G. Pearson[3], R. Gallino[5,6] and C. Zimmermann[1]

[1]Centre de Recherches Pétrographiques et Géochimiques (CRPG), Nancy Université, CNRS, BP 20, 54501 Vandoeuvre-lès-Nancy cedex, France (reisberg@crpg.cnrs-nancy.fr)

[2]Origins Laboratory, Department of the Geophysical Sciences and Enrico Fermi Institute, The University of Chicago, 5734 South Ellis Ave., Chicago IL 60637, USA (dauphas@uchicago.edu)

[3]Department of Earth Sciences, University of Durham, Durham, DH1 3LE, UK (d.g.pearson@durham.ac.uk)

[4]Mineralogisch-Petrologisches Institut, Bonn Universitat, Poppelsdorfer Schloss, Bonn 53115, Germany (ambre.luguet@uni-bonn.de)

[5]Dipartimento di Fisica Generale, Universita' di Torino, via Pietro Giuria 1, 10125 Torino, Italy (gallino@ph.unito.it)

[6]Center for Stellar and Planetary Astrophysics, School of Mathematical Sciences, Monash University, PO Box 28M, Victoria 3800, Australia

**Abstract**

We present osmium isotopic results obtained by sequential leaching of the Murchison meteorite, which reveal the existence of very large internal anomalies of nucleosynthetic origin ($\varepsilon^{184}$Os from -108 to 460; $\varepsilon^{186}$Os from -14.1 to 12.6; $\varepsilon^{188}$Os from -2.6 to 1.6; $\varepsilon^{190}$Os from -1.7 to 1.1). Despite these large variations, the isotopic composition of the total leachable osmium (weighted average of the leachates) is close to that of bulk chondrites. This is consistent with efficient large-scale mixing of Os isotopic anomalies in the protosolar nebula. The Os isotopic anomalies are correlated, and can be explained by the variable contributions of components derived from the *s*, *r* and *p*-processes of nucleosynthesis. Surprisingly, much of the *s*-process rich osmium is released by relatively mild leaching, suggesting the existence of an easily leachable *s*-process rich presolar phase, or alternatively, of a chemically resistant *r*-process rich phase. Taken together with previous evidence for a highly insoluble *s*-process rich carrier, such as SiC, these results argue for the presence of several presolar phases with anomalous nucleosynthetic compositions in the Murchison meteorite. The *s*-process composition of Os released by mild leaching diverges slightly from that released by aggressive digestion techniques, perhaps suggesting that the presolar phases attacked by these differing procedures condensed in different stellar environments. The correlation between $\varepsilon^{190}$Os and $\varepsilon^{188}$Os can be used to constrain the *s*-process $^{190}$Os/$^{188}$Os ratio to be 1.275±0.043. Such a ratio can be reproduced in a nuclear reaction network for a MACS value for $^{190}$Os of ~200 ± 22 mbarn at 30 keV. More generally, these results can help refine predictions of the *s*-process in the Os mass region, which can be used in turn to constrain the amount of cosmoradiogenic $^{187}$Os in the solar system and hence the age of the Galaxy.

We also present evidence for extensive internal variation of $^{184}$Os abundances in the Murchison meteorite. A steep anti-correlation is observed between $\varepsilon^{184}$Os and $\varepsilon^{188}$Os. Since $^{184}$Os is formed uniquely by the *p*-process, this anti-correlation cannot be explained by variable addition or subtraction of *s*-process Os to average solar system material. Instead, this suggests that *p* or *r*-process rich presolar grains (e.g., supernova condensates) may be present in meteorites in sufficient quantities to influence the Os isotopic compositions of the leachates. Nevertheless, $^{184}$Os is a low abundance isotope and we cannot exclude the possibility that the measured anomalies for this isotope reflect unappreciated analytical artifacts.

**Keywords:** nucleosynthetic anomalies; Os isotopes; osmium; Murchison meteorite; *s*-process; *p*-process; carbonaceous chondrites

# 1. Introduction

Primitive, unequilibrated chondritic meteorites provide precious information about the origin of the material from which our solar system formed. These meteorites have not been homogenized by thermal metamorphism. They thus retain large internal isotopic variations that testify to the fine scale heterogeneity of the presolar nebula and place strong constraints on the nucleosynthetic processes that occur in stars. Presolar grains in primitive meteorites often have extreme isotopic compositions that reflect the specific stellar environments in which they formed (see review by Zinner, 1998). Isotopic variations arising from the slow-neutron capture process (*s*-process) of nucleosynthesis can be studied by a variety of methods. These include *in situ* techniques, which allow the isotopic compositions of individual presolar grains to be determined (Barzyk, et al., 2006; Nicolussi, et al., 1997; Nicolussi, et al., 1998a; Nicolussi, et al., 1998b; Savina, et al., 2003; Savina, et al., 2004; Terada, et al., 2006; Zinner, et al., 1991), bulk analysis of concentrates of presolar grains (Ott and Begeman, 1990; Podosek, et al., 2004; Prombo, et al., 1993; Yin and Lee, 2006) and sequential acid leaching of powdered bulk rock (Dauphas, et al., 2002c; Hidaka, et al., 2003; Schönbächler, et al., 2005). These techniques are complementary - the first reveals the extreme and often highly variable isotopic compositions found among presolar grains from a single meteorite, while the second and third can be used to precisely characterize the average isotopic signatures of the main nucleosynthetic components. Previous leaching studies of primitive chondrites revealed the existence of fine-scale anomalies of Mo (Dauphas, et al., 2002c), Ba (Hidaka, et al., 2003), and Zr (Schönbächler, et al., 2005), interpreted to result from the greater or lesser contribution of a nucleosynthetic component formed by the *s*-process. This component is thought to be hosted by presolar phases, dissolved to varying extents by the different leaching steps.

The first study of nucleosynthetic Os anomalies was by Brandon et al. (2005), who analyzed bulk rock powders of various chondritic meteorites. While ordinary and enstatite chondrites displayed no resolvable anomalies, the Tagish Lake carbonaceous chondrite showed significant deficits in Os isotopes manufactured by the *s*-process. These were attributed to incomplete digestion of a highly insoluble presolar phase. This explanation was supported by Yokoyama et al. (2007), who showed that very aggressive digestions of acid leaching residues display excesses in *s*-process Os nuclides. These authors also confirmed the uniform bulk Os isotopic composition (except for $^{187}$Os) of the various chondrite classes after total sample digestion. More recently, Yokoyama et al. (2008) reported large isotopic anomalies in insoluble organic matter extracted from chondrites.

We present Os isotopic results from a sequential leaching study of the Murchison meteorite, a primitive (CM2) chondrite that fell in Victoria, Australia in 1969. This meteorite, which experienced very little thermal metamorphism, retains a high abundance of presolar grains and has been the subject of many studies aimed at understanding stellar nucleosynthesis. Large internal isotopic variations have been observed for several elements including Ba (Hidaka, et al., 2003) and Zr (Schönbächler, et al., 2005). As is true of other elements, Os nucleosynthetic anomalies can provide information about stellar environments of element production and about the distribution of presolar dust grains in the protosolar nebula. In addition, better knowledge of the *s*-process abundance of $^{187}$Os would allow calculation of the cosmoradiogenic $^{187}$Os component, thus placing constraints on the age of the Galaxy (Clayton, 1988; Mosconi, et al., 2006; Yokoi, et al., 1983). The sequential leaching results presented below provide new information on the nature of the presolar phases present in meteorites. Comparison of these and previous results with state-of-the-art modeling of *s*-process nucleosynthesis in AGB stars places strong constraints on the stellar environment in which these nuclides formed.

## 2. Analytical Techniques

The analytical procedure is described in detail in the supplemental on-line material (SOM). In brief, ~16.5 g of powdered Murchison meteorite were subjected, at the University of Chicago, to the following sequential leaching procedure (concentrated acids were used and diluted as indicated):

Step 1)  50 mL acetic acid + 50 mL $H_2O$, 1 day, 20 °C
Step 2)  25 mL $HNO_3$ + 50 mL $H_2O$, 5 days, 20 °C
Step 3)  30 mL HCl + 35 mL $H_2O$, 1 day, 75 °C
Step 4)  30 mL HF + 15 mL HCl + 15 mL $H_2O$, 1 day, 75 °C
Step 5)  10 mL HF + 10 mL HCl, 3 days, 150 °C
Step 6)  2 mL $HNO_3$ + 2 mL HF, 120 °C, 15 hours (applied to a fraction of Step 5 residue)

A fraction of each leachate was removed for Os analysis. Os, Re and Pt separation was performed at the Centre de Recherches Pétrographiques et Géochimiques (CRPG). About 10 to 20% of each Os fraction was used for isotope dilution concentration determinations (after spiking with $^{190}$Os, $^{185}$Re, and $^{196}$Pt), while the rest was used for high precision Os isotopic measurements. Both aliquots of each fraction were oxidized in HCl-$HNO_3$ mixtures in Carius tubes (Shirey and Walker, 1995), except for step 6, which was oxidized with $CrO_3$ in a teflon vessel. After oxidation, Os was extracted into $Br_2$, and purified by microdistillation (Birck, et al., 1997). Re and Pt were extracted from the residual aqueous phase using method 1 of Rehkämper and Halliday (1997). Re and Pt concentrations were determined from isotopic analyses by ICPMS (Elan 6000) at the analytical service (SARM) of CRPG. Total blanks were 2 pg for Re and 46 pg for Pt. Os was analyzed as $OsO_3^-$ by negative thermal ionization mass spectrometry (NTIMS) (Creaser, et al., 1991; Volkening, et al., 1991). Concentrations were calculated from isotopic analyses on the CRPG Finnigan MAT262 mass spectrometer. High precision Os isotopic compositions were determined by multi-faraday cup measurements of the unspiked aliquots on the Durham University Triton mass spectrometer (detailed procedure in Luguet et al. (2008); points of particular interest for this study highlighted in SOM). Following Yokoyama et al. (2007), mass fractionation was corrected by normalization to $^{192}$Os/$^{189}$Os = 2.527411. This isotopic pair is used for normalization because both $^{192}$Os and $^{189}$Os were formed in great majority by the *r*-process (Table 2). For the same reason, isotopic abundances are reported relative to $^{189}$Os, rather than the more commonly used $^{188}$Os, because the *s*-process contribution to $^{189}$Os (~4%) is much smaller than its contribution to $^{188}$Os (~28%). Total Os blanks, from the Carius tube step on, were ~0.5 pg. Os blanks of each leaching step are difficult to characterize, but since purified reagents were used, these were probably insignificant relative to the large quantities of Os in the samples.

The Os standard solution used in Durham was obtained from the University of Maryland (standard UMd) and is the same as that used by both Yokoyama et al. (2007) and Brandon et al. (2005). Results from the three laboratories (Table 1) are generally in good agreement, with similar reproducibility. Os isotopic compositions of the samples are presented in epsilon units, where:

$$\varepsilon^i\text{Os} = 10^4 * [(^i\text{Os}/^{189}\text{Os})/(^i\text{Os}/^{189}\text{Os})_{std}] - 1] \text{ and } i = 184, 186, 187, 188, \text{ or } 190$$

The UMd isotopic ratios determined in Durham were used as the reference values for all isotopes except $^{186}$Os. Since this standard has a radiogenic $^{186}$Os/$^{189}$Os composition, we used instead the average chondritic $^{186}$Os/$^{189}$Os ratio (0.0982524) determined by Yokoyama et al. (2007) to calculate $\varepsilon^{186}$Os, which facilitated comparison with previous studies. Use of the Durham value of the UMd standard ratio would result in $\varepsilon^{186}$Os values ~0.6 units lower than those listed.

## 3. Results

*3.1 Os, Re and Pt concentrations of the various leachate fractions*

Os, Re and Pt contents released by the different leaching steps vary markedly (Fig. 1). The listed concentrations (Table 1) represent the quantity of Os, Re or Pt released by each leaching step, divided by the total mass of the sample. Most of the Os (~ 70%) is contained in the first two fractions. The total Os concentration calculated for the Murchison meteorite (547 ± 18 ppb), obtained by summing the Os released by each leaching step, is 5 to 30% lower than those determined by previous bulk rock analyses (580 to 759 ppb) (Horan, et al., 2003; Jochum, 1996; Walker, et al., 2002a; Walker and Morgan, 1989).

The total calculated Re concentration (46 ± 7 ppb) is 20% lower than the bulk rock values (54.5 and 56.3 ppb) of Jochum (1996) and Walker and Morgan (1989), but in excellent agreement with more recent results (49.5 and 46.0 ppb) of Walker et al. (2002a) and Horan et al. (2003). Re is even more strongly concentrated than Os in the first leachate fraction (Fig. 1). This is not surprising, as Re is more easily released during sample dissolution than Os (Meisel, et al., 2003). When $^{187}$Os/$^{188}$Os is plotted against $^{187}$Re/$^{188}$Os, the bulk Murchison composition (Walker, et al., 2002b) plots on the 4.56 Ga iron meteorite isochron (Smoliar, et al., 1996) (Fig. 2), though this may be fortuitous given the large uncertainty on the $^{187}$Re/$^{188}$Os ratio. In contrast, the leachate fractions are not aligned along this isochron. Walker et al. (2002b) attributed the slightly non-isochronous Re-Os behavior of certain bulk carbonaceous chondrite samples to late-stage open-system behavior. Another, perhaps more likely cause of the dispersion of our leachate data may be incongruent dissolution of the constitutent phases. Certain elements, such as Re, may be released preferentially during the leaching process. In this case, the Re/Os ratios of the leachates would not correspond to those of the mineral phases present in the meteorite, making it impossible to derive an isochron.

The total calculated Pt concentration (900 ± 17 ppb), like that of Os, is 15 to 20% lower than previously reported values (1120, (Jochum, 1996); 1043 ppb, (Horan, et al., 2003)). Our estimated bulk Pt/Os ratio (1.65) is in the range of previous values (1.48 to 1.80). Pt is released less than either Re or Os during the initial leaching steps, but is more strongly concentrated in the final leaching step. This is consistent with the results of Yokoyama et al. (2007) who found that acid-resistant residues of the Murchison, Allende and Tagish Lake meteorites had super-chondritic Pt/Os ratios.

*3.2 Os isotopic compositions*

Os isotopic results are listed in Table 1, and plotted in Fig. 3. The case of $^{184}$Os, a pure *p*-process isotope, will be discussed separately. Except for minor *p*-process and radiogenic contributions, $^{186}$Os is an *s*-only isotope, $^{188}$Os and $^{190}$Os are of mixed *s-r* origin, and $^{189}$Os and $^{192}$Os are almost pure *r*-process nuclides (Table 2). Leachates 2, 3 and 4 show strong enrichments in $\varepsilon^{186}$Os, and moderate enrichments in $\varepsilon^{188}$Os and $\varepsilon^{190}$Os. The spectra observed for these leachates are consistent with an excess of an *s*-process-rich component in these fractions. In contrast, leachates 1 and 5 show strong depletions in $\varepsilon^{186}$Os and moderate depletions in $\varepsilon^{188}$Os and $\varepsilon^{190}$Os, indicative of an *s*-process deficit. These latter spectra are consistent with the *s*-process-poor spectra obtained for the Tagish Lake carbonaceous chondrite (Brandon, et al., 2005). Leachate 6 shows a slight enrichment in *s*-process. (Semantically, *s*-process excess is identical to *r*-process deficit and vice-versa.)

The *s*-process-rich and *s*-process-poor spectra are strikingly complementary. The weighted average of the spectra (Fig. 3b) was obtained by multiplying the measured epsilon values by the fraction of Os in each leachate. The weighted average $\varepsilon^{188}$Os and $\varepsilon^{190}$Os values are indistinguishable from zero, while that of $\varepsilon^{186}$Os is just slightly higher than zero. Thus despite the internal heterogeneity revealed by the leachates, the reconstructed bulk Os isotopic composition of Murchison is almost equivalent to that of the silicate earth and of most chondrites.

The $\varepsilon^{186}$Os composition, which is most strongly affected by *s*-process subtraction or addition, is shown as a function of leachate strength in Fig. 4. While the weakest leachate (1) is depleted in the *s*-process component, leachates 2, 3, and 4 are strongly enriched in this component. As leachates 2 and 3 do not contain HF, which attacks silicate phases, this implies that most of the *s*-process-rich Os is released without extensive silicate destruction, though digestion of fine-grained olivine and leaching of other silicates could occur. Leachate 5 is depleted in the *s*-process component, while leachate 6, the most aggressive leaching step, displays minor *s*-process enrichment.

An excellent correlation, including the Murchison leachates from our study, the carbonaceous chondrite residues and Carius tube (CT) digestions from Yokoyama et al. (2007) and the bulk CT analyses of the Tagish Lake chondrite (Brandon, et al., 2005), is observed between

$\varepsilon^{190}$Os and $\varepsilon^{188}$Os (Fig. 5a). The line shown in this figure is not a regression line but rather the correlation expected for the addition or subtraction of Os derived from the *s*-process to Os of terrestrial composition. The *s*-process composition is based on our calculation of *s*-process nucleosynthesis, using the neutron capture cross sections of Bao et al. (2000) and the AGB modelling parameters of the FRANEC models used by Zinner et al. (2006). Dauphas et al. (2004) derived the equation that governs the mixing line in Fig. 5 for internally normalized ratios. For Os, this takes the form:

$$\varepsilon^{i}\text{Os} = \frac{\rho^{i}_{\text{Os}} - \rho^{192}_{\text{Os}}\mu^{i}_{\text{Os}}}{\rho^{188}_{\text{Os}} - \rho^{192}_{\text{Os}}\mu^{188}_{\text{Os}}} \varepsilon^{188}\text{Os} \qquad (1)$$

The parameter $\rho^{i}_{\text{Os}}$ represents the *s*-process composition normalized to the average solar system (*ss*) osmium isotopic composition derived from bulk chondrite measurements, which should be equivalent to the average terrestrial composition.

$$\rho^{i}_{\text{Os}} = \frac{{}^{i}\text{Os}/{}^{189}\text{Os}_{s}}{{}^{i}\text{Os}/{}^{189}\text{Os}_{ss}} - 1 \qquad (2)$$

The parameter $\mu^{i}_{\text{Os}}$ is the mass difference relative to that of the normalizing pair ($^{192}$Os and $^{189}$Os).

$$\mu^{i}_{\text{Os}} = \frac{i - 189}{192 - 189} \qquad (3)$$

A correlation also exists between $\varepsilon^{186}$Os and $\varepsilon^{188}$Os (Fig. 5b), though it is less well-defined than that between $\varepsilon^{190}$Os and $\varepsilon^{188}$Os. Two predicted correlation lines are shown for reference, based on different estimates of the composition of the *s*-process component. The more recent estimate (Table 2) takes into account new measurements of the Maxwellian Averaged Cross-Sections (MACS) of $^{186}$Os, $^{187}$Os, and $^{188}$Os (Mosconi, et al., 2006). Leachates 1, 5, and 6, and the Tagish Lake bulk analyses, have Os compositions consistent with the Mosconi et al. estimated composition, while leachates 2, 4 and especially 3 plot above this line. Yokoyama et al. (2008) also noted departure from a single mixing line in a leachate of Murchison insoluble organic matter, which they attributed to the presence of a presolar carrier of *r*-process Os of unusual composition.

The $^{184}$Os results must be treated with caution, because of the very low abundance of this isotope and the necessity for large corrections for isobaric interferences from PtO$_2$ and WO$_3$ (see SOM). Nevertheless, the good agreement between the preliminary and final analyses of leachate 1, despite their very different signal intensities and interference corrections (8 and 4%, respectively), suggests that the data are reliable. A strong anti-correlation is observed between $\varepsilon^{184}$Os and $\varepsilon^{188}$Os (Fig. 5c). Leachate 2 plots slightly below this trend, but this may reflect the exceptionally large (~10%), and probably overestimated (see SOM), interference correction for this sample. Unlike other Os isotopes, $^{184}$Os was formed uniquely by the *p*-process. Thus addition or subtraction of an *s*-process component has very little effect on $\varepsilon^{184}$Os, resulting in a nearly horizontal array in this diagram (slope = -0.37). Instead, the data define a steep negative trend (slope = -132 ±34), that cannot be explained simply by variable inclusion of Os derived from the *s*-process. This trend does not pass through the origin (y-intercept = 100 ±38 (2σ), excluding leachate 2), unlike the correlations involving the other Os isotopes. The reconstructed bulk $\varepsilon^{184}$Os composition is 34 ±103 (including leachate 2), and is thus consistent with either a chondritic or a moderately super-chondritic bulk composition (the UMd standard used for normalization is assumed to have a chondritic $^{184}$Os composition). Previous bulk (Carius tube) analyses of chondrites (Brandon, et al., 2005) yielded $\varepsilon^{184}$Os values close to zero (Fig. 5c), though statistically significant differences were found between separate fractions of individual meteorites (Allende, Tagish Lake).

## 4. Discussion

*4.1. Nature of the phase enriched in s-process osmium*

The extreme internal osmium isotopic variation revealed by the leaching experiments indicates that the Murchison meteorite contains material formed in diverse nucleosynthetic environments. With the exception of $^{184}$Os, this heterogeneity can be explained by the variable contribution of presolar phases enriched in Os generated by the *s*- or *r*-processes. The exact nature of this phase (or phases) is difficult to constrain. Several studies have shown that presolar SiC grains are often strongly enriched in *s*-process nuclides, including those of osmium (Yokoyama, et al., 2007), molybdenum (Nicolussi, et al., 1998a) and zirconium (Nicolussi, et al., 1997). However, SiC is a highly insoluble phase. In our study, large enrichments in *s*-process osmium were observed in leaching steps (2 and 3) performed at moderate temperatures and without HF (Fig. 4). These steps are unlikely to release osmium contained within the structure of SiC grains, unless these are extremely fine-grained and contain appreciable quantities of Os. Instead, our results suggest that a large fraction of the *s*-process osmium is contained within more soluble presolar phases. Conversely, it is possible that the *s*-process deficits observed in leachates 1 and 5 reflect dissolution of a currently unidentified presolar phase rich in *r*-process nuclides.

Summation of the osmium in the leachate fractions suggests an overall Os concentration (547 ± 18 ppb; Table 1) slightly lower than that obtained by direct bulk rock analyses of Murchison (580 - 759 ppb) (Horan, et al., 2003; Jochum, 1996; Walker, et al., 2002a; Walker and Morgan, 1989). Minor heterogeneity in platinoid contents within Murchison could explain this low total Os content, but this possibility would be rather fortuitous. Volatile Os loss during leaching is an unlikely explanation, as the total Pt content is also low. However, we cannot completely exclude Os volatilization, as Pt recoveries could be low for other reasons, leading by chance to a chondritic bulk Pt/Os ratio. Perhaps the simplest hypothesis is that a small fraction of the Os in the meteorite was not dissolved by any of the leaching steps. Support for this idea comes from the fact that the reconstituted $^{187}$Os/$^{188}$Os ratio (0.12801 ± 0.00012; Table 1) is higher than those obtained from previous bulk analyses of Murchison and other carbonaceous chondrites (0.1262 ± 0.0006, Walker et al., 2002a), suggesting the existence of an undissolved fraction with a complementary, low

$^{187}$Os/$^{188}$Os ratio. The weighted average compositions of the other Os isotopes are indistinguishable from ($\epsilon^{188}$Os, $\epsilon^{190}$Os), or slightly higher than ($\epsilon^{186}$Os), those of most chondrites, including Murchison (Brandon, et al., 2005; Yokoyama, et al., 2007) (Fig. 3b). Thus any Os not dissolved during the leaching procedure must also have, on average, an approximately chondritic composition, or perhaps a minor depletion in *s*-process Os. This finding seems to contradict the results of Brandon et al. (2005) and of Yokoyama et al (2007), who dissolved several carbonaceous chondrites, including Murchison, by the Carius tube (CT) technique. They found that several of the resulting solutions showed deficits in *s*-process Os, suggesting that any undissolved fraction would be enriched in *s*-process-rich components. While CT dissolution does not involve HF, it is an aggressive procedure (concentrated HNO$_3$-HCl at 240 $^o$C for several days). Thus it seems surprising that the CT technique produces solutions with an *s*-process deficit, while the weighted average of our leachates has a chondritic composition, or even a slight *s*-process excess. The last three steps of our leaching procedure employ HF, which could conceivably attack certain phases, such as SiC, not accessed by the CT technique. However, the weighted average of these three steps ($\epsilon^{186}$Os = – 1.4; $\epsilon^{188}$Os = – 0.2; $\epsilon^{189}$Os = – 0.2) has an *s*-process depleted composition, inconsistent with the *s*-process-rich composition required for the residue of the CT leachates. Thus it is unlikely that these leaching steps significantly attacked the phases that were not digested by CT. Yokoyama et al. (2007) leached Murchison following procedures (Lewis, et al., 1975; Ott, et al., 1981) that differed somewhat from ours, and analyzed the residues. The residues digested by highly aggressive techniques, including combustion at high temperature under oxygen, displayed strong *s*-process enrichments. This apparent contradiction with our results could be resolved if any undissolved residue left by our leaching procedure included both *s*- and *r*-process-rich components, despite having an overall nearly chondritic composition.

Taken together, our study and those of Brandon et al. and Yokoyama et al. suggest that osmium nucleosynthetic anomalies could be carried by several distinct presolar phases. One is a highly insoluble phase such as SiC, revealed by the earlier studies. The nature of the others are not clearly defined but could include an easily leachable phase enriched in the *s*-process that is revealed by our leaching steps 2, 3, and 4. Conversely, we cannot exclude the existence of a chemically resistant phase enriched in the *r*-process that was not dissolved in those steps. In their study of Zr isotopes, Schönbächler et al. (2005) concluded that at least two presolar phases, SiC and a more soluble phase, are required to explain the observed nucleosynthetic anomalies. The same scenario might be proposed to explain the *s*-process-rich molydenum obtained during leaching of the Orgueil meteorite (Dauphas, et al., 2002c). Though the authors of this study favored the hypothesis that this anomalous Mo was derived from leaching of SiC grains, it may instead have resulted from dissolution of a more soluble presolar phase. Thus the nucleosynthetic anomalies of osmium and perhaps other elements could be explained by the presence of more than one presolar phase. Whether these phases are part of the inventory of presolar grains already identified remains to be determined.

*4.2. Towards a better definition of the s-process nucleosynthetic composition*

The observed correlations of $\varepsilon^{186}$Os and $\varepsilon^{190}$Os with $\varepsilon^{188}$Os are consistent with the variable addition or subtraction of *s*-process Os to solar (i.e. terrestrial) composition. The exact correlation lines expected depend on the precise composition of the *s*-process (Eq. 1). Brandon et al. (2005) noted a small difference between the measured $\varepsilon^{186}$Os compositions of the Tagish Lake chondrite and those expected from their calculated solar *s*-process mixing trend. To explain this subtle difference, they postulated that the *s*-process component of the Tagish Lake meteorite formed in a stellar environment with a neutron density about twice that of the environment that produced solar *s*-process osmium. However, the *s*-process estimates they used to derive the solar mixing trend were not based on the most recent nucleosynthetic measurements and modeling. Fig. 5b shows mixing trends based on two different *s*-process compositions calculated using state-of-the art nuclear reaction network and AGB-star modeling. One is similar to the compositions used by Brandon et al. (2005) and is based on the nucleosynthetic parameters of Bao et al. (2000), while the other is based on more recent measurements of the MACS of $^{186}$Os, $^{187}$Os, and $^{188}$Os (Mosconi et al., (2006). The MACS is a measure of the capacity of a nucleus to capture neutrons during the *s*-process (Fig. 6). The new results indicate a value about ~25% smaller than the old value for the $^{188}$Os MACS, implying that it is more difficult to transform $^{188}$Os into $^{189}$Os than previously believed. The new *s*-process $^{188}$Os abundance is therefore higher than that obtained by previous calculations.

Both the Tagish Lake bulk rock analyses and our results from leachates 1, 5, and 6, fall close to the correlation line predicted from the Mosconi et al. (2006) results (Fig. 5b). Thus it is not necessary to invoke *s*-process nucleosynthesis in a high neutron flux environment (Brandon, et al., 2005), as these data can be explained by use of the more recent MACS $^{188}$Os to predict the *s*-process Os abundances (Humayun and Brandon, 2007; Reisberg, et al., 2007). One complication is that recent studies have shown that the neutron capture cross section of $^{185}$W is smaller than previously thought (Mohr, et al., 2004; Sonnabend, et al., 2003). This would lead to increased *s*-process flow through $^{185}$Re (Fig. 6) and ultimately to $^{186}$Os abundances (Table 2) substantially higher than the observed solar system value (Meyer and Wang, 2007). Meyer and Wang suggested that this problem could be resolved if the MACS $^{186}$Os value were 20% larger than believed. However, the new result by Mosconi et al. only slightly modified the previous MACS $^{186}$Os value. An alternative solution proposed by Meyer and Wang (2007) might be to increase the neutron capture flow across $^{186}$Re, resulting in the bypass of $^{186}$Os. However this seems unlikely because, at stellar temperatures in He thermal pulse conditions in AGB stars, the importance of a dedicated treatment of the isomeric level of $^{186}$Re is severely reduced by the thermal population of the many low-lying excited states (Franz Käppeler, personal communication). A possible resolution of this problem may be obtained from examination of all of the uncertainties. Both the Os cosmic abundance determined from CI chondrites (uncertainty ~ 8%: (Anders and Grevesse, 1989; Lodders, 2003) and the theoretical $^{186}$Os production rates are poorly constrained. Uncertainties on the latter value include those on the MACS $^{186}$Os value, the experimental MACS of unstable $^{185}$W (up to ~ 30%; (Bao, et al., 2000; Mohr, et al., 2004; Sonnabend, et al., 2003)), the stellar enhancement factor (~5%;

(Holmes, et al., 1976; Rauscher and Thielemann, 2000) and the β decay meanlife of $^{185}$W at stellar conditions (up to ~30%, (Goriely, 1999), based on the theoretical expectations of (Takahashi and Yokoi, 1987). Finally, the average theoretical reproducibility of all other heavy *s*-only isotopes belonging to the main *s*-component must be considered. Thus the most probable explanation of the apparent $^{186}$Os overproduction problem is that it is an artifact resulting from incomplete consideration of the uncertainties.

Leachates 2, 3, and possibly 4, which are also leachates rich in *s*-process osmium, have $\varepsilon^{186}$Os compositions that plot above the *r-s* mixing line derived from the Mosconi et al. results (Fig. 5b). In particular, leachate 3 plots five epsilon units above this line. Potential isobaric interferences were carefully measured before and after each analysis, and were much too small to explain an offset of this magnitude. As we can think of no other analytical artifact that could cause this discrepancy, we suspect that the $^{186}$Os excesses of these leachates result from a natural process. Several possibilities can be examined: 1) *Radiogenic ingrowth of $^{186}$Os*. Since $^{186}$Os is the daughter of the very long-lived isotope $^{190}$Pt ($t_{1/2}$ ~480 Ga), it could be suggested that radioactive decay produced some of the observed variation in $\varepsilon^{186}$Os. In particular, if the phases dissolved by leachates 2 and 3 had much higher Pt/Os ratios than the other leachates, their $^{186}$Os excesses could result from radiogenic ingrowth. Correction of the leachate and the chondritic $^{186}$Os/$^{189}$Os ratios for 4.56 Ga of radioactive decay using the measured Pt/Os ratios (e.g., Brandon et al. (2005) and Yokoyama et al. (2007), see SOM) has little effect on the calculated $\varepsilon^{186}$Os anomalies (Fig. 7). However, by analogy with the observed effects on Re/Os ratios, the measured Pt/Os ratios may have been modified by open system behavior and/or the leaching procedure, making them inappropriate to use for radiogenic corrections. In theory, the Os in leachate 3 could have been originally associated with a phase with an extremely high Pt/Os ratio, thus explaining its elevated $\varepsilon^{186}$Os value. However this would require association of nearly all of the Pt in the meteorite with only the Os in this leachate, an adhoc and implausible suggestion, which would not leave enough Pt left over to explain the high $\varepsilon^{186}$Os values of steps 2 and 4. Thus radiogenic ingrowth is unlikely to explain the $^{186}$Os excesses. Finally, we note that while a vague correlation (not shown) exists between $^{186}$Os/$^{189}$Os and $^{190}$Pt/$^{189}$Os if leachate 6 is excluded, this trend would yield an age of ~70 Ga, and is thus clearly without time significance. 2) *Galactic cosmic rays*. Interaction with galactic cosmic rays (GCR) could be proposed to explain the observed $^{186}$Os excesses. However, this seems very unlikely as the Murchison meteorite has a low exposure age. Although cosmogenic effects on Os isotopes have been noted in iron meteorites (Huang and Humayun, 2008), in general chondrites do not show effects associated with GCR irradiation. 3) *p-process contribution*. Since $^{186}$Os includes a small component (~2-3%) of *p*-process origin, the $^{186}$Os excesses of leachates 2 and 3 could reflect excesses of *p*-process Os in these phases. Osmium-184, a pure *p*-process isotope, provides a direct constraint on the contribution of *p*-process anomalies to $^{186}$Os. Isotopic anomalies arising from variations in the *p*-process alone should correlate according to:

$$\varepsilon^{186}_{Os} = \frac{^{186}Os/^{184}Os_p}{^{186}Os/^{184}Os_{ss}} \varepsilon^{184}_{Os} \qquad (4)$$

Using the values of Table 2, we predict that the anomalies on $^{186}$Os from the *p*-process should be 0.024 times those on $^{184}$Os, with a factor of ~2 uncertainty on the proportionality constant. The effect of correcting $\varepsilon^{186}$Os for the possible presence of *p*-process anomalies using $\varepsilon^{184}$Os is shown in Fig. 7. For leachates 1 and 5, the correction leads to a significant negative shift in $\varepsilon^{186}$Os and improves the correlation between $\varepsilon^{188}$Os and $\varepsilon^{186}$Os. However, the slope of this correlation (9.4) is much steeper than that predicted from the MACS measured by Mosconi et al. For leachates 2, 3 and 4, the correction, though small, is positive and thus increases their deviation from the predicted trend. 4) *Variable s-process composition.* As noted above, when combined with previous studies, our work suggests the presence of two distinct presolar hosts of *s*- or *r*-process Os, one of which may be an easily leachable phase, while the other is highly insoluble. The *s*-rich, highly insoluble Murchison residues of Yokoyama et al. (2007) plot close to the Mosconi et al. mixing line (Fig. 5b), while our leachates 2 and 3 plot above this line. This might suggest that the insoluble and leachable phases contain *s*-process Os of slightly different compositions.

Thus we tentatively suggest that we see evidence for two different *s*-process compositions in the Murchison leachates, hosted by different phases. Assuming that the *s*-process enrichment of fractions 2, 3 and 4 results from the dissolution of a highly soluble *s*-process rich phase, rather than the non-dissolution of an *r*-process rich phase, these distinct *s*-process compositions might ultimately be linked to changes in the C/O ratio during the evolution of AGB stars (Lodders and Amari (2005) and references therein). At the beginning of the *s*-process, the C/O ratio is low, and primarily oxides and silicates are expected to form. As the C/O ratio increases, SiC and graphite can condense. The *s*-process composition associated with these 2 stages is not expected to remain constant because the $^{13}$C($\alpha$,n) and $^{22}$Ne($\alpha$,n) neutron sources are activated to different extents. In particular, in the more advanced episodes at C/O>1, the $^{22}$Ne($\alpha$,n)$^{25}$Mg reaction provides a neutron burst with a peak neutron density that increases with the thermal pulse number. This increases the effect of the neutron branching channel at $^{191}$Os (Fig. 6), which will decrease $^{192}$Os *s*-process abundances. While this nuclide has only a minor *s*-process component, it is an isotope that is used for internal normalization and can thus affect all of the ratios. Taking all third dredge up episodes together (when convection of the envelope penetrates into the He intershell, carrying newly synthesized matter to the surface of the AGB star), the predicted slope of the $\varepsilon^{186}$Os vs $\varepsilon^{188}$Os correlation is 4.92. (This is the slope expected for mainstream presolar SiC, and therefore differs from the slope of ~4.4 in Fig. 5b, which corresponds to the *s*-process composition of the solar system obtained using the modelling parameters of Arlandini et al., 1999.) If one separates the *s*-process yields for third dredge up episodes with C/O>1 or C/O<1, the slopes are 4.89 and 5.10, respectively. These slopes were calculated by summing over the lifetime of the AGB-star all osmium ejected in winds when the composition of the envelope is C/O>1 or C/O<1, respectively. Oxide/silicate presolar phases (C/O<1) would be expected to be more easily leached and have higher slopes. SiC (C/O>1) would be expected to be more resistant to acid leaching and have lower slopes. This is consistent with what is observed (steps 2 and 3 have higher slopes while the acid-resistant residues analyzed by Yokoyama et al. (2007) define a lower slope).

*4.3 Estimation of the s-process $^{190}Os/^{188}Os$ ratio*

Mosconi et al. (2006) did not determine new MACS for $^{189}$Os and $^{190}$Os, so their new data cannot be used to calculate the *s*-process mixing trend in Fig. 5a. On the other hand, as suggested by Humayun and Brandon (2007), the observed mixing trend may be used to constrain the *s*-process yields for $^{190}$Os. Based on the slope of the trend of the Brandon et al. (2005) data in the plot of $\varepsilon^{190}$Os vs. $\varepsilon^{188}$Os (0.587 ± 0.058), these authors proposed a value of 249 ± 36 (all uncertainties 2σ) mbarns for the neutron capure cross section of $^{190}$Os ($\sigma_{190}$). In Fig. 8 we repeat this exercise, using all of the data now available (this study; (Brandon, et al., 2005; Yokoyama, et al., 2008; Yokoyama, et al., 2007). We find a slope (*m*) of 0.651 ± 0.028, which is related to the *s*-process $^{190}Os/^{188}Os$ ratio by the following equation (derivation in SOM):

$$\left(\frac{^{190}Os}{^{188}Os}\right)_s = \left(\frac{^{190}Os}{^{188}Os}\right)_{ss} \left(m + R^{189}(1 - \mu^{190} - m + m\mu^{188}) + R^{192}(\mu^{190} - m\mu^{188})\right) \quad (5)$$

where $R^i_{Os} = \frac{(^iOs/^{188}Os)s}{(^iOs/^{188}Os)ss}$

and $\mu^i_{Os}$ is defined in equation (3). Equation (5) is somewhat different from that of Humayun and Brandon, because it includes terms that account for the *s*-process components of $^{189}$Os and $^{192}$Os. Based on the data in Table 2, the ratio ($^{190}Os/^{188}Os$)$ss$ is 1.983. Using the modelling results discussed above, we find that $R^{189}(1-\mu^{190}-m+m\mu^{188}) = -0.0307$ and $R^{192}(\mu^{190} - m\mu^{188}) = 0.0225$. This translates into an *s*-process $^{190}Os/^{188}Os$ ratio of 1.275±0.043.

To obtain their estimate of $\sigma_{190}$, Humayun and Brandon assumed that the local approximation applied (abundance × cross section=constant). However, our calculations show that the local approximation is not valid for $^{188}$Os and $^{190}$Os (Fig. 9). This stems from the fact that $^{188}$Os is fed by several branching points, in particular $^{185}$W (Fig. 6). The predicted $^{190}Os/^{188}Os$ *s*-process ratio was therefore computed using a nuclear reaction network for different values of $\sigma_{190}$, assuming a fixed value of 291 ±30 mbarns for $\sigma_{188}$ (Mosconi et al. 2006). As was expected, the $^{190}Os/^{188}Os$ ratio scales as $1/\sigma_{190}$ (Fig. 9). The estimated *s*-process $^{190}Os/^{188}Os$ ratio of 1.275 ±0.043 can be reproduced using a cross section of $\sigma_{190}$ = 200 ±22 mbarn (includes uncertainty on $\sigma_{188}$). This is much lower than the value of 295 ±45 mbarn reported by Bao et al. (2000). There are several other ways to explain the *s*-process $^{190}Os/^{188}Os$ ratio of 1.27. For instance, increasing the nuclear flow from $^{185}$W towards $^{185}$Re can increase the $^{190}Os/^{188}Os$ ratio but this would create a more severe overproduction problem for $^{186}$Os.

*4.4 Implications for the distribution of s-process carrier phases in the presolar nebula*

The leachate data presented above strengthen the evidence for large internal Os isotopic anomalies of nucleosynthetic origin in chondritic meteorites. Nevertheless, bulk chondrite analyses, after dissolution by techniques assuring total sample digestion, indicate surprising Os isotopic homogeneity both within and between the various chondrite classes (Yokoyama, et al., 2007), suggesting that on a large scale *s*-process and *r*-process components were very well mixed in the presolar nebula. This conclusion is consistent with Zr isotopic data from bulk chondrites, eucrites, lunar and terrestrial samples (Schönbächler, et al., 2003) and with certain studies of Mo and Ru isotopes in chondrites (Becker and Walker, 2003a; Becker and Walker, 2003b). On the other hand, other studies of Mo and Ru (Chen, et al., 2003; Chen, et al., 2004; Dauphas, et al., 2002a; Papanastassiou, et al., 2004; Yin, et al., 2002) and of Ba and Nd isotopes (Andreasen and Sharma, 2006; Carlson, et al., 2007; Hidaka, et al., 2003; Ranen and Jacobsen, 2006) provide evidence of small but measurable *p*-, *s*- and *r*-process anomalies in bulk chondrites. Some of these apparent anomalies may result from incomplete sample digestion, as demonstrated for Os isotopes (Yokoyama et al. (2007); Brandon et al. (2005)). However, all of the evidence for isotopic variations on large spatial scales cannot be dismissed on this basis. For example, *p*-process and probable *s*-process anomalies in Sm and Nd were determined for the Allende meteorite after alkali fusion, a total digestion procedure (Carlson, et al., 2007). The slight *s*-process deficiency they observed for Nd is consistent with the *s*-process deficiency in Mo isotopes found in bulk and all leachate analyses of this meteorite (Dauphas, et al., 2002c; Yin, et al., 2002). In addition, *s*-process deficits observed in iron meteorites (Chen, et al., 2003; Chen, et al., 2004; Dauphas, et al., 2002b; Papanastassiou, et al., 2004; Qin, et al., 2008) cannot be ascribed to incomplete sample digestion, as iron meteorites, which form by the fusion and differentiation of large volumes of solar system material, should have internally homogeneous isotopic compositions. Furthermore, iron meteorite Mo and Ru anomalies are correlated, and plot along the trend predicted for the mixing of *s*-process with *p* and *r*-process components (Dauphas, et al., 2004). Thus different elements and different studies provide conflicting answers concerning the degree of large-scale heterogeneity in the presolar nebula.

*4.5 Os derived from the p-process*

As shown in Fig. 5c, a strong anti-correlation exists between $\varepsilon^{184}$Os and $\varepsilon^{188}$Os, implying that the proportion of Os derived from the *p*-process increases markedly as that from the *s*-process decreases. Since $^{184}$Os is a *p*-only isotope, this anti-correlation is much too steep to be explained by variable addition of *s*-process Os to material with a constant *p*-process/*r*-process ratio. Instead, the *p/r* ratio is apparently highest where the *r*-process contribution is highest, and the *s*-process contribution is lowest. Thus the data define a rough linear mixing array between an endmember with a high p/r and low s/r ratio, and another with a low p/r ratio and high s/r ratio. Such an array could be produced by the variable but correlated contribution of two distinct presolar phases, carrying Os generated by different nucleosynthetic processes. These could be *s+r*, *r+p* or *s+p*; mathematically it is not possible to distinguish between these possibilities. The bulk chondrite analyses of Brandon et al. (2005) do not plot on the correlation line in Fig. 5c, but instead cluster about the x-axis. This may be because the internal nucleosynthetic anomalies observed in our study were carried by leachable phases; in contrast, the Brandon et al. (2005) study would likely reveal only those anomalies carried by highly insoluble phases, such as SiC. Further work is warranted to test whether the $^{184}$Os anomalies are real, and if so, to identify their possible presolar carrier(s).

The reconstituted bulk $\varepsilon^{184}$Os composition, obtained from the weighted average of the leachates, is 34 ±103 (Table 1, Fig. 3). However, this value may be biased by the large contribution of leachate 2, which might be influenced by an analytical artifact. The anti-correlation in Fig. 5c does not pass through the origin, but instead crosses the y-axis at 100 ± 38 $\varepsilon^{184}$Os units (excluding leachate 2). This suggests that the bulk leachable composition of Murchison may in fact be moderately enriched in *p*-process Os. This suggestion should be considered with prudence, given the large uncertainties in the $^{184}$Os data. Furthermore, the bulk leachable composition may not be equivalent to the true bulk composition, since it is likely that some highly insoluble phases have not been digested. This suggestion is supported by the data of Brandon et al. (2005), who determined $\varepsilon^{184}$Os values close to zero for 13 chondrites digested by CT. On the other hand, these authors also found statistically distinct $\varepsilon^{184}$Os values for different powder aliquots of the Allende (-24±5 vs. 45±5, relative to the UMd standard) and Tagish Lake (from -31±16 to 23±15) meteorites, providing supporting evidence for the existence of internal heterogeneities. In constrast with our Os results, bulk analyses of both the Murchison and the Allende meteorites display deficits in the *p*-only isotope $^{144}$Sm (Andreasen and Sharma, 2006; Carlson, et al., 2007). Thus, if a bulk enrichment in $^{184}$Os in Murchison is confirmed by future studies, it will be necessary to explain how enrichment of *p*-process nuclides of some elements can be coupled with *p*-process depletions of other elements within the same meteorite.

## 5. Conclusions

Sequential leaching of the Murchison meteorite revealed the presence of large internal osmium isotopic anomalies of nucleosynthetic origin. Despite these variations, the weighted average of the leachate compositions is close to that of bulk chondrites. The observed isotopic anomalies are strongly inter-correlated, and are consistent with mixing, in different proportions in the various leachates, of components derived from the slow (*s*), rapid (*r*), and proton (*p*) nucleosynthetic processes. The $\varepsilon^{190}$Os vs. $\varepsilon^{188}$Os correlation, including all of the currently available data (this study; Brandon, et al., 2005; Yokoyama, et al., 2008; Yokoyama, et al., 2007) and the most recent MACS value for $^{188}$Os (Mosconi et al., (2006), were used to estimate the *s*-process $^{190}$Os/$^{188}$Os (1.275 ±0.043). To explain this value, the MACS of $^{190}$Os may need to be revised to 200 ±22 mbarn.

Relatively mild leaching procedures released Os with a strong excess in *s*-process (or deficit in *r*-process). This suggests the existence of at least two presolar carriers of anomalous osmium in the Murchison meteorite: one, revealed by this study, is an unidentified phase (*s*-process rich and easily leachable or *r*-process rich and chemically resistant), while the other, revealed by the work of Brandon et al. (2005), and Yokomaya et al. (2007), is a highly insoluble phase, most likely SiC. A small difference exists between the *s*-process compositions revealed by mild (our study) and highly aggressive (previous studies) digestion techniques. Assuming that we are indeed seeing evidence for the existence of an easily leachable *s*-process rich presolar phase, the *s*-process compositions of these two carrier phases may parallel changes in the C/O ratio of the stellar environment in which the grains condensed.

This study also reports data suggesting large internal abundance variations of $^{184}$Os, an isotope formed uniquely by the *p*-process. A strong anti-correlation exists between $\varepsilon^{184}$Os and $\varepsilon^{188}$Os that cannot be explained by variable addition of an *s*-process component to material with a constant *p*-process/*r*-process ratio. Instead this trend, coupled with the correlations between $\varepsilon^{186}$Os, $\varepsilon^{188}$Os and $\varepsilon^{190}$Os, suggests the variable but correlated contributions of two presolar phases which carry Os derived from distinct nucleosynthetic processes, at least one of which must be the *r* or more likely the *p* process. However, $^{184}$Os is a low abundance isotope so we cannot exclude the possibility that the measured anomalies reflect unappreciated analytical artifacts. More work is clearly warranted to elucidate the origin of these variations.

**Acknowledgements**

We thank Munir Humayun and an anonymous EPSL reviewer for their thoughtful and constructive comments. We also thank Andrew M. Davis, Meenakshi Wadhwa, Roy S. Lewis and Robert N. Clayton for discussions and support, and Marita Mosconi and Franz Käppeler for much advice. The Murchison specimen used for the leachate study was generously provided by the Field Museum. This work was supported by grants from the French INSU program PNP (to LR), the National Aeronautics and Space Administration (grant NNG06GG75G), the France-Chicago Center, and the Packard Fellowship program (to ND) and the Italian MIUR-PRIN06 Project "Late phases of Stellar Evolution: Nucleosynthesis in Supernovae, AGB stars, Planetary Nebulae" (to RG). AL thanks the European Union for the Marie Curie Post-doctoral Fellowship (EIF-ENV-009752) during which this study was performed. This is CRPG contribution number XXXX.

**References**

Anders, E. and Grevesse, N., 1989. Abundances of the elements: Meteoritic and solar, Geochim. Cosmochim. Acta 53, 197-214.

Andreasen, R. and Sharma, M., 2006. Solar nebular heterogeneity in *p*-process samarium and neodymium isotopes, Science 314, 806-809.

Bao, Z.Y., Beer, H., Käppeler, F., Voss, F. and Wisshak, K., 2000. Neutron cross sections for nucleosynthesis studies, At. Data Nucl. Data Tables 76, 70-154.

Barzyk, J.G., Savina, M.R., Davis, A.M., Gallino, R., Pellin, M.J., Lewis, R.S., Amari, S. and Clayton, R.N., 2006. Multi-element isotopic analysis of single presolar SiC grains, New Astron. Rev. 50, 587-590.

Becker, H. and Walker, R.J., 2003a. Efficient mixing of the solar nebula from uniform Mo isotopic composition of meteorites, Nature 425, 152-155.

Becker, H. and Walker, R.J., 2003b. In search of extant Tc in the early solar system: 98Ru and 99Ru abundances in iron meteorites and chondrites, Chem. Geol. 196, 43-56.

Birck, J.-L., Roy Barman, M. and Capmas, F., 1997. Re-Os isotopic measurements at the femtomole level in natural samples, Geostandards Newsletter 21, 19-27.

Brandon, A.D., Humayun, M., Puchtel, I.S., Leya, I. and Zolensky, M., 2005. Osmium isotope evidence for an s-process carrier in primitive chondrites, Science 309, 1233-1236.

Carlson, R.W., Boyet, M. and Horan, M., 2007. Chondrite barium, neodymium, and samarium isotopic heterogeneity and early earth differentiation, Science 316, 1175-1178.

Chen, J.H., Papanastassiou, D.A. and Wasserburg, G.J., 2003. Endemic Ru isotopic anomalies in iron meteorites and in Allende, Lunar Planet. Sci. XXXIV, 1789.

Chen, J.H., Papanastassiou, D.A. and Wasserburg, G.J., 2004. Endemic Mo isotopic anomalies in iron and carbonaceous meteorites, Lunar Planet. Sci. XXXV, 1431.

Clayton, D.D., 1988. Nuclear cosmochronology within analytic models of the chemical evolution of the solar neighbourhood, Mon. Not. R. Astron. Soc. 234, 1-36.

Creaser, R.A., Papanastassiou, D. and Wasserburg, G.J., 1991. Negative thermal ion mass spectrometry of osmium, rhenium, and iridium, Geochim. Cosmochim. Acta 55, 397-401.

Dauphas, N., Marty, B., Davis, A.M. and Reisberg, L., 2004. The cosmic molybdenum–ruthenium isotope correlation, Earth Planet. Sci. Lett. 226, 465-475.

Dauphas, N., Marty, B. and Reisberg, L., 2002a. Inference on terrestrial genetics from molybdenum isotope systematics, Geophys. Res. Lett., Geophys. Res. Lett. 29(6), 409-419.

Dauphas, N., Marty, B. and Reisberg, L., 2002b. Molybdenum evidence for inherited planetary scale isotope heterogeneity of the protosolar nebula, Astrophys. J. 565, 640-644.

Dauphas, N., Marty, B. and Reisberg, L., 2002c. Molybdenum nucleosynthetic dichotomy revealed in primitive meteorites, Astrophys. J. 569, L139-L142.

Goriely, S., 1999. Uncertainties in the solar system r-abundance distribution, Astron. Astrophys. 342, 881-891.

Hidaka, H., Ohta, Y. and Yoneda, S., 2003. Nuclosynthetic components of the early solar system inferred from Ba isotopic compositions in carbonaceous chondrites, Earth Planet. Sci. Lett. 214, 455-466.

Holmes, J.A., Woosley, S.E., Fowler, W.A. and Zimmerman, B.A., 1976. Tables of Thermonuclear-Reaction-Rate Data for Neutron-Induced Reactions on Heavy Nuclei, Atomic Data Nuclear Data Tables 18, 305-412.

Horan, M.F., Walker, R.J., Morgan, J.W., Grossman, J.N. and Rubin, A.E., 2003. Highly siderophile elements in chondrites, Chem. Geol. 196, 5-20.

Huang, S. and Humayun, M., 2008. Osmium isotope anomalies in group IVB irons: Cosmogenic or nucleosynthetic contributions, in: LPSC XXXIX, pp. #1168, Houston.

Humayun, M. and Brandon, A.D., 2007. *S*-process implications from osmium isotope anomalies in chondrites, Astrophys. J. 664, L59-L62.

Jochum, K.P., 1996. Rhodium and other platinum-group elements in carbonaceous chondrites, Geochim. Cosmochim. Acta 60, 3353-3357.

Lewis, R.S., Srinivasan, B. and Anders, E., 1975. Host phase of a strange xenon component in Allende, Science 190, 1251-1262.

Lodders, K., 2003. Solar system abundances and condensation temperatures of the elements, Astrophys. J. 591, 1220-1247.

Lodders, K. and Amari, S., 2005. Presolar grains from meteorites: Remnants from the early times of the solar system, Chemie der Erde 65, 93-166.

Ludwig, K.R., 2003. Isoplot 3.00 A Geochronological Toolkit for Microsoft Excel, Berkeley Geochronology Center.

Luguet, A., Nowell, G.M. and Pearson, D.G., 2008. $^{184}$Os/$^{188}$Os and $^{186}$Os/$^{188}$Os measurements by negative thermal ionisation mass spectrometry (NTIMS): Effects of interfering element and mass fractionation corrections on data accuracy and precision, Chem. Geol. 248 (special issue HSE Geochemistry), 342-362.

Meisel, T., Reisberg, L., Moser, J., Carignan, J., Melcher, F. and Brügmann, G., 2003. Re-Os systematics of UB-N, a serpentinized peridotite reference material, Chem. Geol. 201, 161-179.

Meyer, B.S. and Wang, C., 2007. *S*-process branching at $^{186}$Re, the abundance of $^{186}$Os, and presolar grains, in: Lunar Planet. Sci. XXXVIII, pp. 2055, Houston.

Mohr, P., Shizuma, T., Ueda, H., Goko, S., Makinaga, A., Hara, K.Y., Hayakawa, T., Lui, Y.-W., Ohgaki, H. and Utsunomiya, H., 2004. *s*-process branching at 185W revised, Phys. Rev. C - Nuclear Physics 69, 032801-1.

Mosconi, M., Heil, M., Käppeler, F., Plag, R., Mengoni, A., Fujii, K., Gallino, R., Aerts, G., Terlizzi, R.et al., 2006. Experimental challenges for the Re/Os clock, Proc. of Science (NIC-IX), #055.

Nicolussi, G.K., Davis, A.M., Pellin, M.J., Lewis, R.S. and Clayton, R.N., 1997. *S*-process zirconium in presolar silicon carbide grains, Science 277, 1281-1283.

Nicolussi, G.K., Pellin, M.J., Lewis, R.S., Davis, A.M., Amari, S. and Clayton, R.N., 1998a. Molybdenum isotopic composition of individual presolar silicon carbide grains from the Murchison metorite, Geochim. Cosmochim. Acta 62, 1093-1104.

Nicolussi, G.K., Pellin, M.J., Lewis, R.S., Davis, A.M., Clayton, R.N. and Amari, S., 1998b. Strontium isotopic composition in individual circumstellar silicon carbide grains: A record of s-process nucleosynthesis, Phys. Rev. Lett. 81.

Ott, U. and Begeman, F., 1990. Discovery of *s*-process barium in the Murchison meteorite, Astrophys. J. 353, L57-L60.

Ott, U., Mack, R. and Chang, S., 1981. Noble-gas-rich separates from the Allende meteorite, Geochim. Cosmochim. Acta 45, 1751-1788.

Papanastassiou, D.A., Chen, J.H. and Wasserburg, G.J., 2004. More on Ru endemic isotope anomalies in meteorites, Lunar Planet. Sci. XXXV, #1828.

Podosek, F.A., Prombo, C.A., Amari, S. and Lewis, R.S., 2004. *s*-process Sr isotopic compositions in presolar SiC from the Murchison meteorite, Astrophys. J. 605, 960-965.

Prombo, C.A., Podosek, F.A., Amari, S. and Lewis, R.S., 1993. *S*-process Ba isotopic compositions in presolar SiC from the Murchison meteorite, Astrophys. J. 410, 393-399.

Qin, L., Dauphas, N., Wadhwa, M., Markowski, A., Gallino, R., Janney, P.E. and Bouman, C., 2008. Tungsten nuclear anomalies in planetesimal cores, Astrophys. J. 674, 1234-1241.

Ranen, M.C. and Jacobsen, S.B., 2006. Barium isotopes in chondritic meteorites: Implications for planetary reservoir models, Science 314, 809-812.

Rauscher, T. and Thielemann, F.-K., 2000. Astrophysical Reaction Rates From Statistical Model Calculations, Atomic Data Nuclear Data Tables 75, 1-351.

Rehkämper, M. and Halliday, A.N., 1997. Development and application of new ion-exchange techniques for the separation of the platinum group and other siderophile elements from geological samples, Talanta 44, 663-672.

Reisberg, L.C., Dauphas, N., Luguet, A., Pearson, D.G. and Gallino, R., 2007. Large *s*-process and mirror osmium isotopic anomalies within the Murchison meteorite, in: LPSC XXXVIII, pp. #1177.

Savina, M.R., Davis, A.M., Tripa, C.E., Pellin, M.J., Clayton, R.N., Lewis, R.S., Amari, S., Gallino, R. and Lugaro, M., 2003. Barium isotopes in individual presolar silicon carbide grains from the Murchison meteorite, Geochim. Cosmochim. Acta 67, 3201-3214.

Savina, M.R., Davis, A.M., Tripa, C.E., Pellin, M.J., Gallino, R., Lewis, R.S. and Amari, S., 2004. Extinct Technetium in Silicon Carbide Stardust Grains: Implications for Stellar Nucleosynthesis, Science 303, 649-652.

Schönbächler, M., Lee, D.-C., Rehkämper, M., Halliday, A.N., Fehr, M.A., Hattendorf, B. and Günther, D., 2003. Zirconium isotope evidence for incomplete admixing of *r*-process components in the solar nebula, Earth Planet. Sci. Lett. 216, 467-481.

Schönbächler, M., Rehkämper, M., Fehr, M.A., Halliday, A.N., Hattendorf, B. and Günther, D., 2005. Nucleosynthetic zirconium isotope anomalies in acid leachates of carbonaceous chondrites, Geochim. Cosmochim. Acta 69, 5113-5122.

Shirey, S.B. and Walker, R.J., 1995. Carius tube digestion for low-blank rhenium-osmium analysis, Anal. Chem. 67, 2136-2141.

Smoliar, M.I., Walker, R.J. and Morgan, J.W., 1996. Re-Os ages of Group IIA, IIIA, IVA, and IVB iron meteorites, Science 271, 1099-1102.

Sonnabend, K., Mohr, P., Vogt, K., Zilges, A., Mengoni, A., Rauscher, T., Beer, H., Käppeler, F. and Gallino, R., 2003. The *s*-process branching at $^{185}$W, Astrophys. J. 583, 506-513.

Takahashi, K. and Yokoi, K., 1987. Beta-decay rates of highly ionized heavy atoms in stellar interiors, Atomic Data Nuclear Data Tables 36, 375-409.

Terada, K., Yoshida, T., Iwamoto, N., Aoki, W. and Williams, I.S., 2006. Eu isotopic analysis of SiC grains from the Murchison Meteorite, in: International Symposium on Origin of Matter and Evolution of Galaxies 2005 847, pp. 324-329, AIP Conf. Proc., 847.

Volkening, J., Walczyk, T. and Heumann, K.G., 1991. Osmium isotope determinations by negative thermal ionization mass spectrometry, Int. J. Mass Spec. Ion Phys. 105, 147-159.

Walker, R.J., Horan, M.F., Morgan, J.W., Becker, H., Grossman, J.N. and Rubin, A.E., 2002a. Comparative $^{187}$Re-$^{187}$Os systematics of chondrites: Implications regarding early solar system processes, Geochim. Cosmochim. Acta 66, 4187-4201.

Walker, R.J. and Morgan, J.W., 1989. Rhenium-osmium isotope systematics of carbonaceous chondrites, Science 243, 519-522.

Walker, R.J., Prichard, H.M., Ishiwatari, A. and Pimentel, M., 2002b. The osmium isotopic composition of the convecting upper mantle deduced from ophiolite chromites, Geochim. Cosmochim. Acta 66, 329-345.

Yin, Q., Jacobsen, S.B. and Yamashita, K., 2002. Diverse supernova sources of pre-solar material inferred from molybdenum isotopes in meteorites, Nature 415, 881-883.

Yin, Q. and Lee, C.-T.A., 2006. Signatures of the *s*-process in presolar silicon carbide grains: barium through hafnium, Astrophys. J. 647, 676-684.

Yokoi, K., Takahashi, K. and Arnould, M., 1983. The $^{187}$Re-$^{187}$Os chronology and chemical evolution of the Galaxy, Astron. Astrophys. 117, 65-82.

Yokoyama, T., Alexander, C.M.O'D. and Walker, R.J., 2008. Osmium isotopic anomalies of insoluble organic matter in chondrites, in: LPSC XXXIX, pp. #1376.

Yokoyama, T., Rai, V.K., Alexander, C.M.O'D., Lewis, R.S., Carlson, R.W., Shirey, S.B., Thiemens, M.H. and Walker, R.J., 2007. Osmium isotope evidence for uniform distribution of *s*- and *r*-process components in the early solar system, Earth Planet. Sci. Lett. 259, 567-580.

Zinner, E., 1998. Stellar nucleosynthesis and the isotopic composition of presolar grains from primitive meteorites, Annu. Rev. Earth Planet. Sci. 26, 147-188.

Zinner, E., Amari, S. and Lewis, R.S., 1991. *s*-process Ba, Nd, and Sm in presolar SiC from the Murchison meteorite, Astrophys. J. 382, L47-L50.

Zinner, E., Nittler, L.R., Gallino, R., Karakas, A.I., Lugaro, M., Straniero, O. and Lattanzio, J.C., 2006. Silicon and Carbon Isotopic Ratios in AGB Stars: SiC Grain Data Models, and the Galactic Evolution of the Si Isotopes, Astrophys. J. 650, 350-373.

**Figure Captions**

**Figure 1.** Pie diagrams showing the distribution of Os, Re and Pt among the leachates. Leaching steps: 1) acetic acid, 20 °C, 1 day; 2) $HNO_3$ + $H_2O$, 20 °C, 5 days; 3) HCl + $H_2O$, 75 °C, 1 day; 4) HF + HCl, 75 °C, 1 day; 5) HF + HCl, 150 °C, 3 days; 6) $HNO_3$ + HF, 120 °C, 15 hours.

**Figure 2.** Re-Os isochron diagram showing the various leachates. A 4.56 Ga isochron (Smoliar, et al., 1996) and bulk analyses of carbonaceous chondrites (small black squares; (Walker, et al., 2002a)) are shown for reference. Open circles represent results of leaching steps, indicated by number; filled blue circle represents the reconstituted bulk Murchison composition calculated from the leachates; open squares represent bulk analyses of Murchison from Walker et al, (2002a). While the bulk composition plots on the isochron, the individual leachates do not, suggesting that the Re/Os systematics have been perturbed by incongruent dissolution during leaching.

**Figure 3.** a) Osmium isotopic compositions in epsilon units as a function of mass number. Mirror anomalies are observed. Leachates 2, 3 and 4 show excesses in an *s*-process rich component, while leachates 1 and 5 show deficits in this component. Leachate 6 displays a small $^{186}Os$ excess. Bulk rock results from the Tagish Lake chondrite (Brandon, et al., 2005) are shown for reference. b) Isotopic spectrum of the weighted average of the leachates. The reconstituted bulk rock anomalies are small ($\varepsilon^{186}Os$) or indistinguishable from zero ($\varepsilon^{184}Os$, $\varepsilon^{188}Os$ and $\varepsilon^{190}Os$) despite the strong variability of the individual leachates.

**Figure 4.** Os derived from the *s*-process, as exemplified by the $\varepsilon^{186}Os$ composition, as a function of increasing leachate strength. Osmium enriched in an *s*-process component is extracted during relatively mild leaching steps.

**Figure 5**. a) $\varepsilon^{190}Os$ vs. $\varepsilon^{188}Os$; b) $\varepsilon^{186}Os$ vs. $\varepsilon^{188}Os$ and c) $\varepsilon^{184}Os$ vs. $\varepsilon^{188}Os$. Numbers indicate leachate fractions. Bulk carbonaceous and enstatite (open circles) and ordinary (open triangles) chondrite data from Brandon, et al., (2005) are shown for comparison. Also shown are bulk analyses and various fractions (Carius tube dissolution, aggressive digestion of acid-leached residues) of carbonaceous chondrites from the study of Yokoyama et al. (2007) (open diamonds). All of the fractions and bulk rock analyses plot along well-defined correlations, which are consistent with the variable contribution of a component enriched in *s*-process osmium. Two predicted trends for addition or subtraction of an *s*-process component are shown for reference. One (dotted line, slope = 7.46) is based on the MACS of Bao et al. (2000), while the other (solid line, slope = 4.37) is based on the much more precise and accurate MACS for $^{186}Os$, $^{187}Os$, and $^{188}Os$

of Mosconi et al. (2006). As these latter authors did not remeasure MACS for $^{189}$Os or $^{190}$Os, only the trend based on the Bao et al. MACS (slope = 0.62) is shown in the plot of $\varepsilon^{190}$Os vs. $\varepsilon^{188}$Os. In (c), the thick gray line represents a linear regression through all the points except leachate 2, while the dashed curved lines represent the 2σ error envelope, calculated using the Isoplot program of Ludwig (2003). Inclusion of leachate 2 would yield a correlation with a slope of -142±77(2σ), a y-intercept of 77±81(2σ) and a MSWD of 2.6. The predicted trend for s-process addition-subtraction in (c) is nearly coincident with the x-axis (slope = -0.37). $\varepsilon^{184}$Os data were taken from the supplemental online material of Brandon, et al. (2005), after renormalization to $^{189}$Os and recalculation relative to their mean $^{184}$Os/$^{189}$Os value for the UMd standard.

**Figure 6**. Chart of the nuclides in the Os region. The s-process path is shown by black arrows; r-process is shown by thick gray arrows. Radioactive decay of the long-lived isotope $^{187}$Re contributes largely to $^{187}$Os (rg denotes radiogenic). The p-process, as well as α decay of the minor isotope $^{190}$Pt, contribute slightly to $^{186}$Os. Short-lived isotopes are shown in italics with half-lives noted. Branching points are observed at $^{185}$W and at $^{186}$Re.

**Figure 7.** Open circles: Effect of $^{190}$Pt decay correction on the $\varepsilon^{186}$Os vs. $\varepsilon^{188}$Os correlation. Leachate $^{186}$Os/$^{188}$Os ratios were corrected for 4.56 Ga of Pt decay using measured Pt/Os ratios and compared to the initial chondritic value. Except for leachate 6, the effects of radiogenic ingrowth are negligible. Green diamonds: Correction of the p-process contribution to $\varepsilon^{186}$Os calculated based on $\varepsilon^{184}$Os variations. Leachates 1 and 5 are the most affected and the correction is significant. However, the correlation becomes steeper (corrected slope = 9.8 ±3.5; uncorrected slope = 6.3 ±2.1), thus even more inconsistent with the MACS measurements of Mosconi et al (2006) (predicted slope = 4.37). The p-process correction of leachates 2, 3 and 4 shifts these points further from the predicted correlation line. Error bars indicate the uncertainty resulting from the p-process corrrection.

**Figure 8.** Plot of $\varepsilon^{190}$Os vs. $\varepsilon^{188}$Os including all of the existing data. (Black crosses: Published data from this study, Brandon et al. (2005), Yokoyama et al. (2007)); Gray crosses: Data from Yokoyama et al. (2008), currently available only in abstract format.) Regression of all data (Isoplot program, K. Ludwig, 2003) yields a correlation line with a slope of 0.651 ± 0.014 (1σ). The corresponding s-process $^{190}$Os/$^{188}$Os ratio is 1.275 ±0.043 (see text, Eq. 5).

**Figure 9.** S-process $^{190}$Os/$^{188}$Os ratio as a function of $^{190}$Os MACS (see text for details). The horizontal line represents the estimated s-process $^{190}$Os/$^{188}$Os ratio (1.275 ±0.043, derived from the slope of the $\varepsilon^{190}$Os vs $\varepsilon^{188}$Os correlation, Eq.5). The local approximation curve, which assumes σ*abundance=constant, was calculated following Humayun and Brandon (2007), using $\sigma_{188}$=291 mbarn (Mosconi et al. 2006). A full nuclear reaction network calculation shows that the local approximation is not valid for $^{188}$Os and $^{190}$Os. Using a realistic model of the s-process the estimated s-process $^{190}$Os/$^{188}$Os ratio of 1.275 can only be reproduced with $\sigma_{190}$ = 200 ±22 mbarn (2σ).

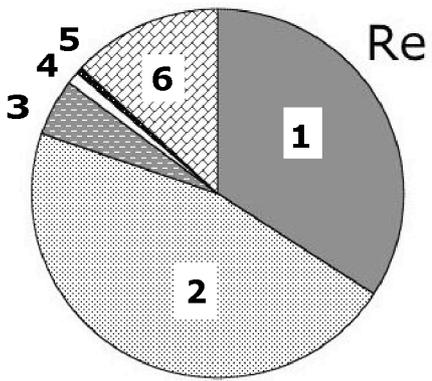 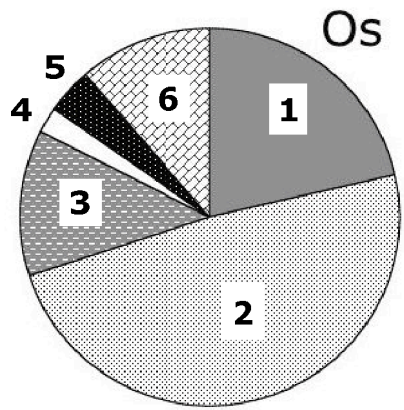 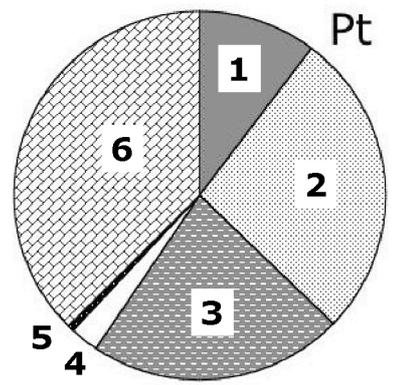

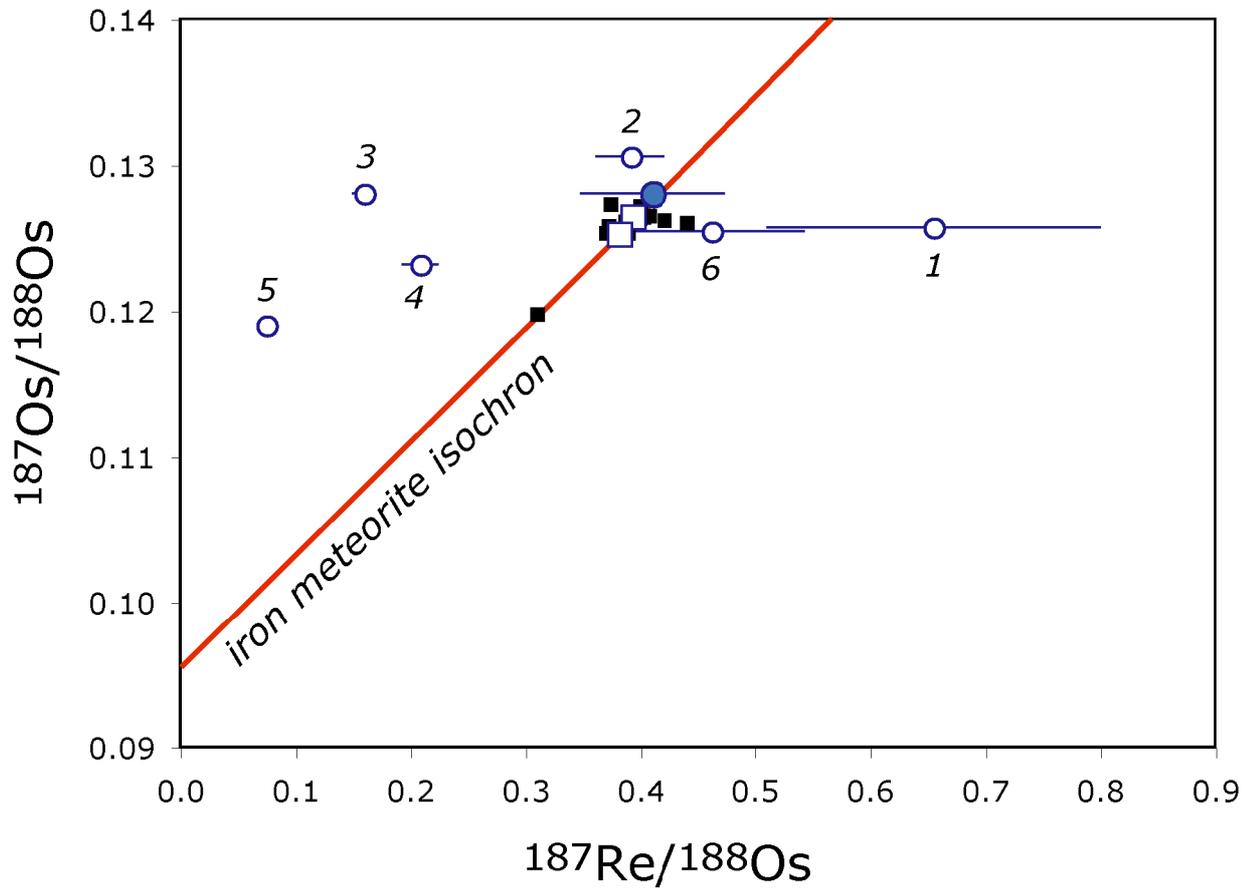

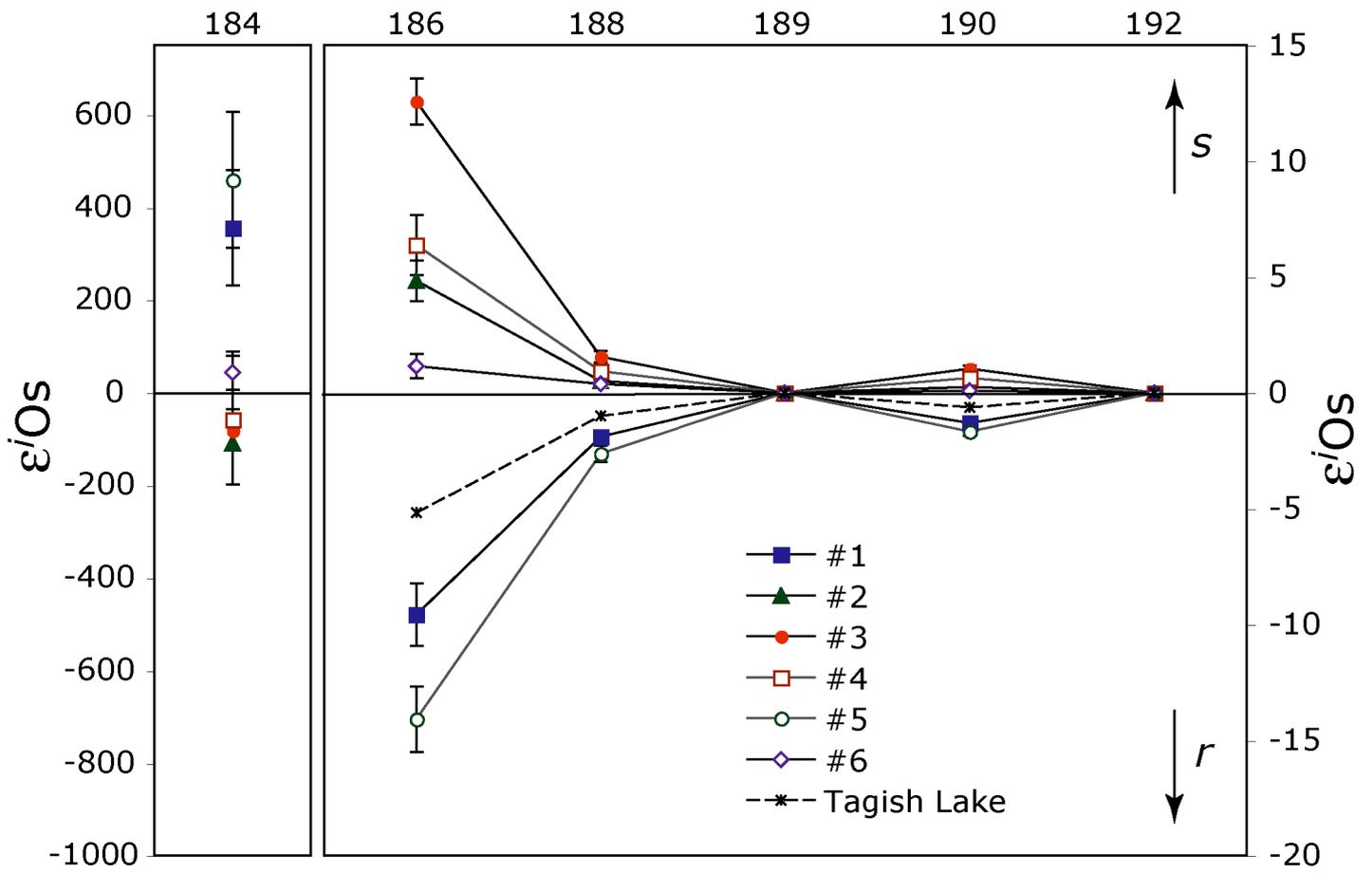
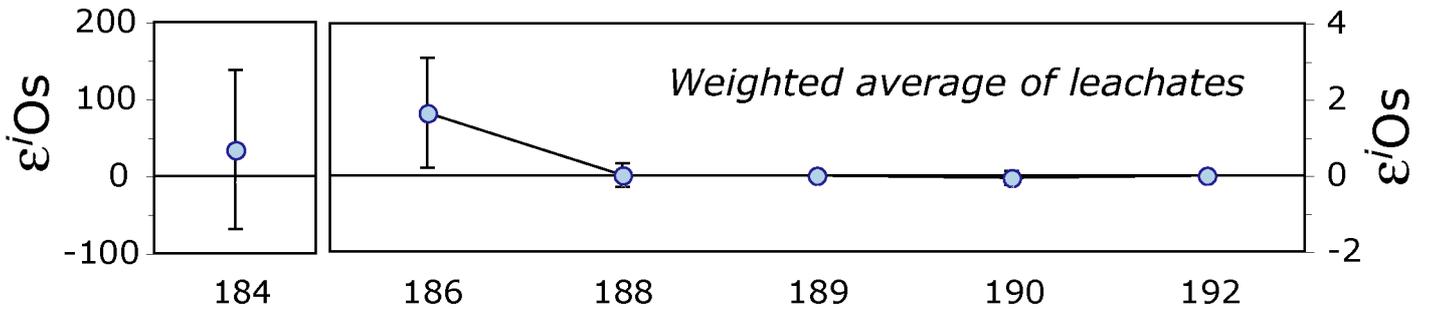

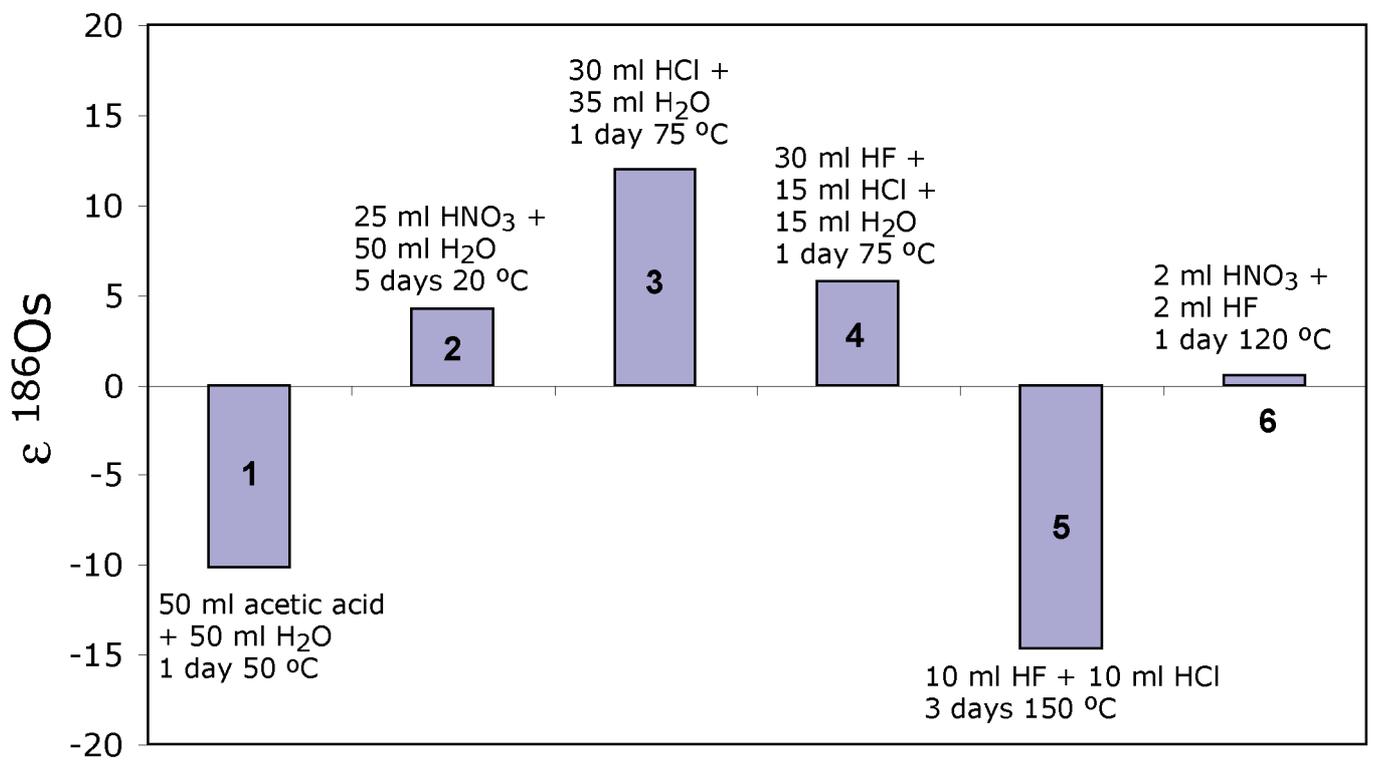

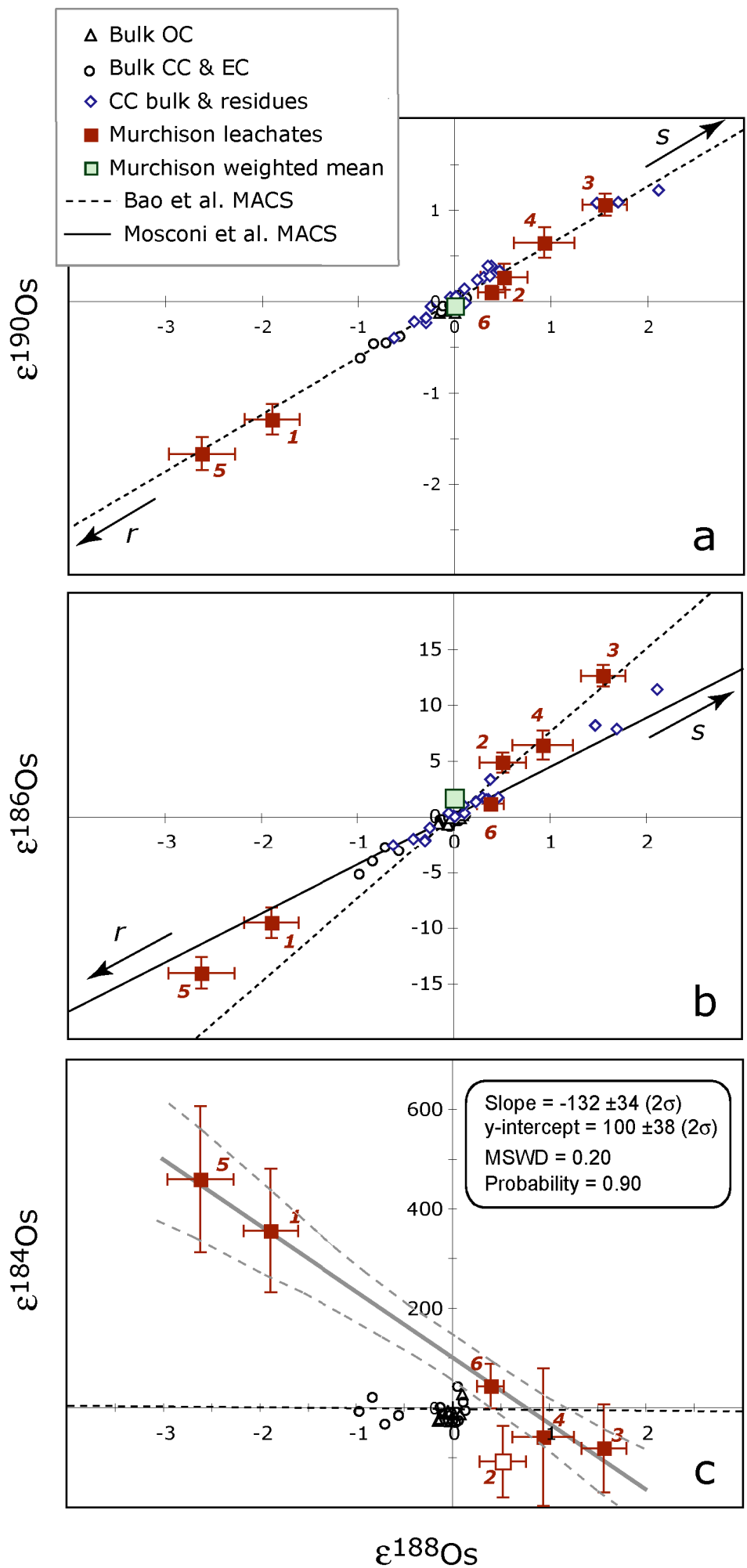

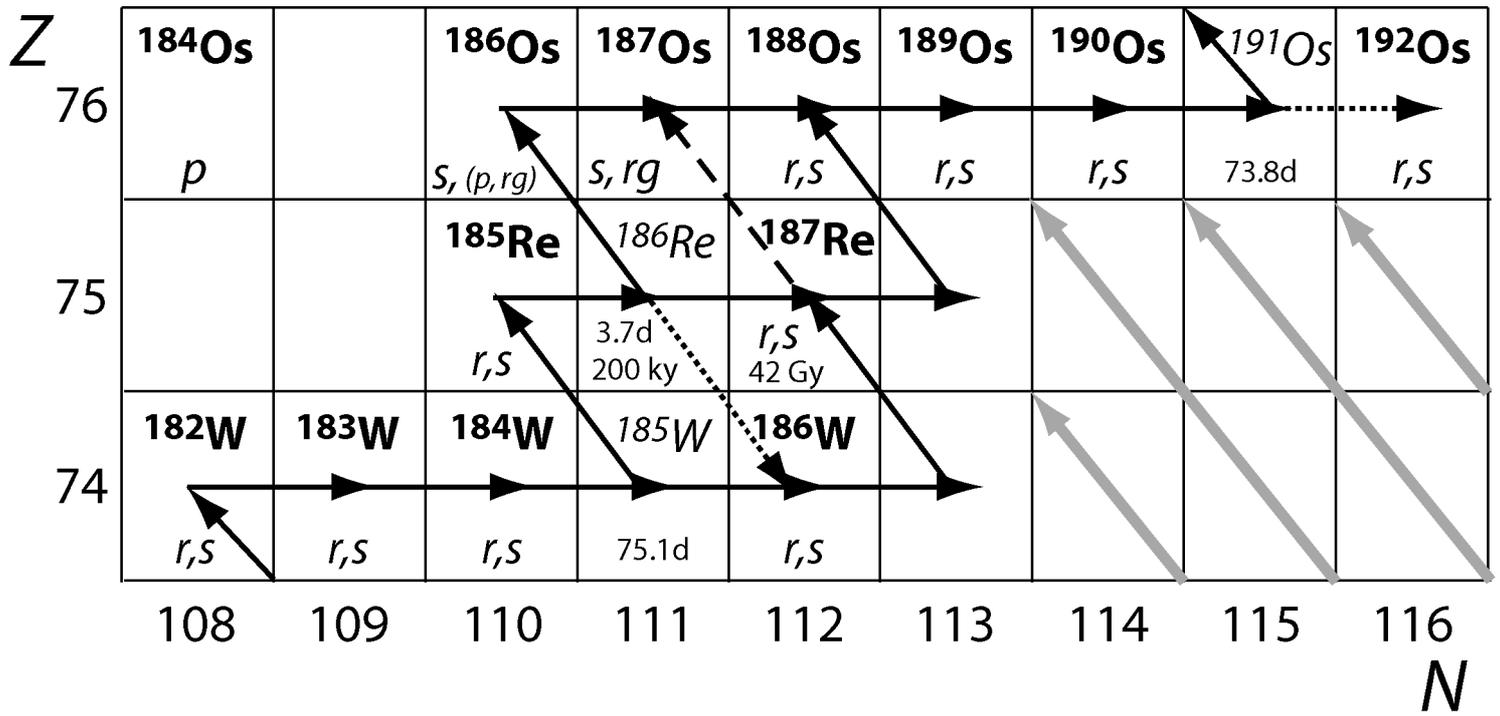

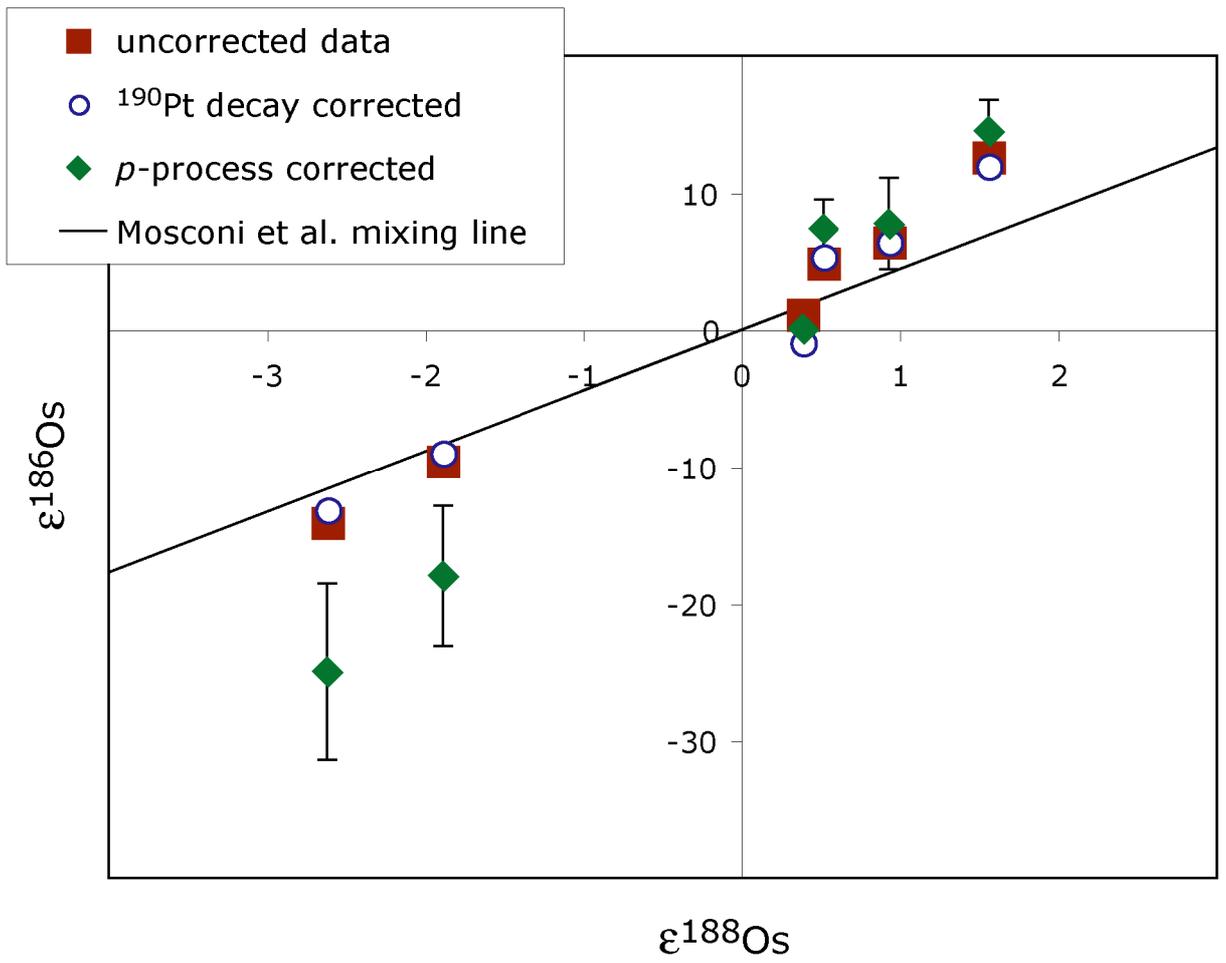

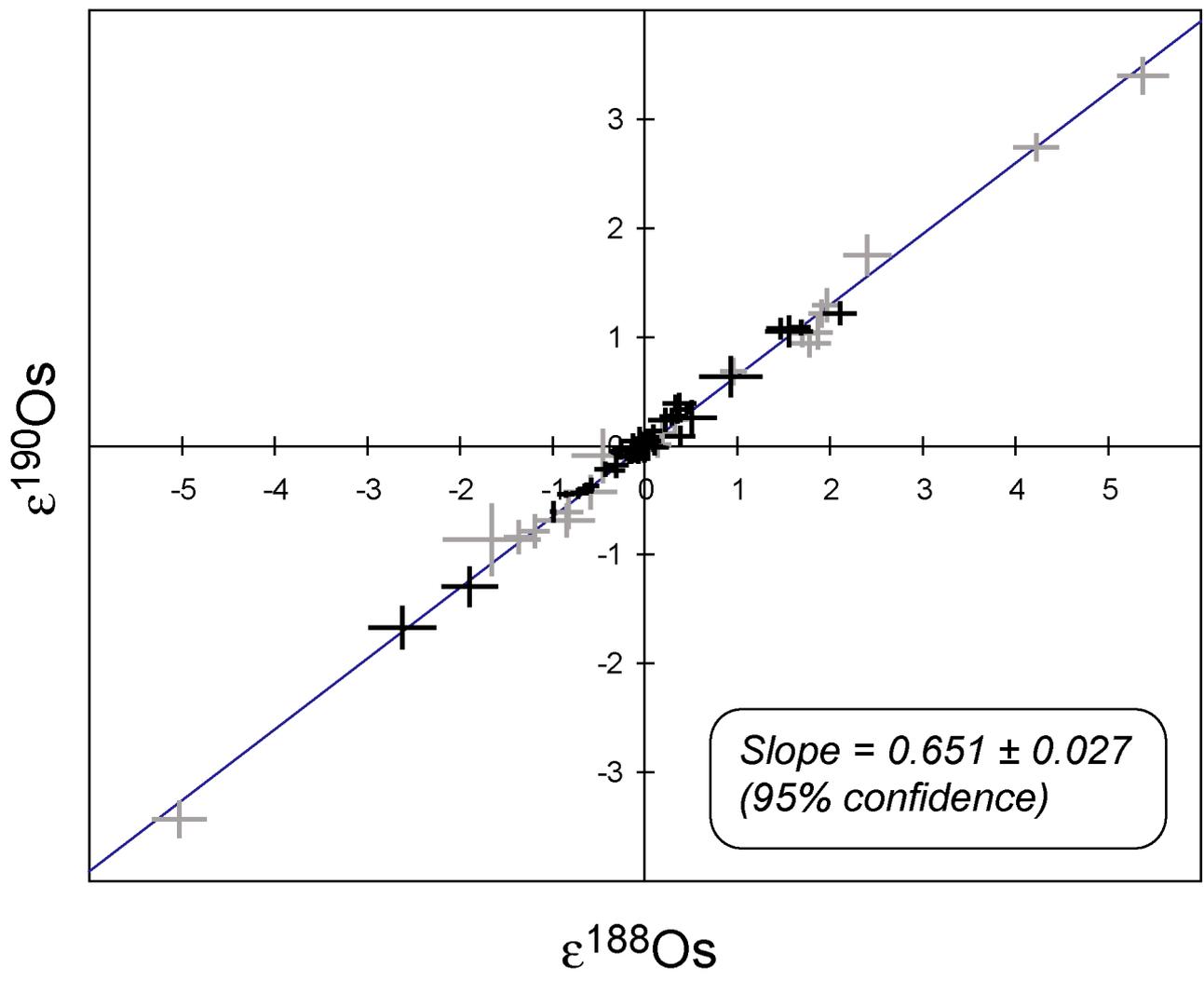

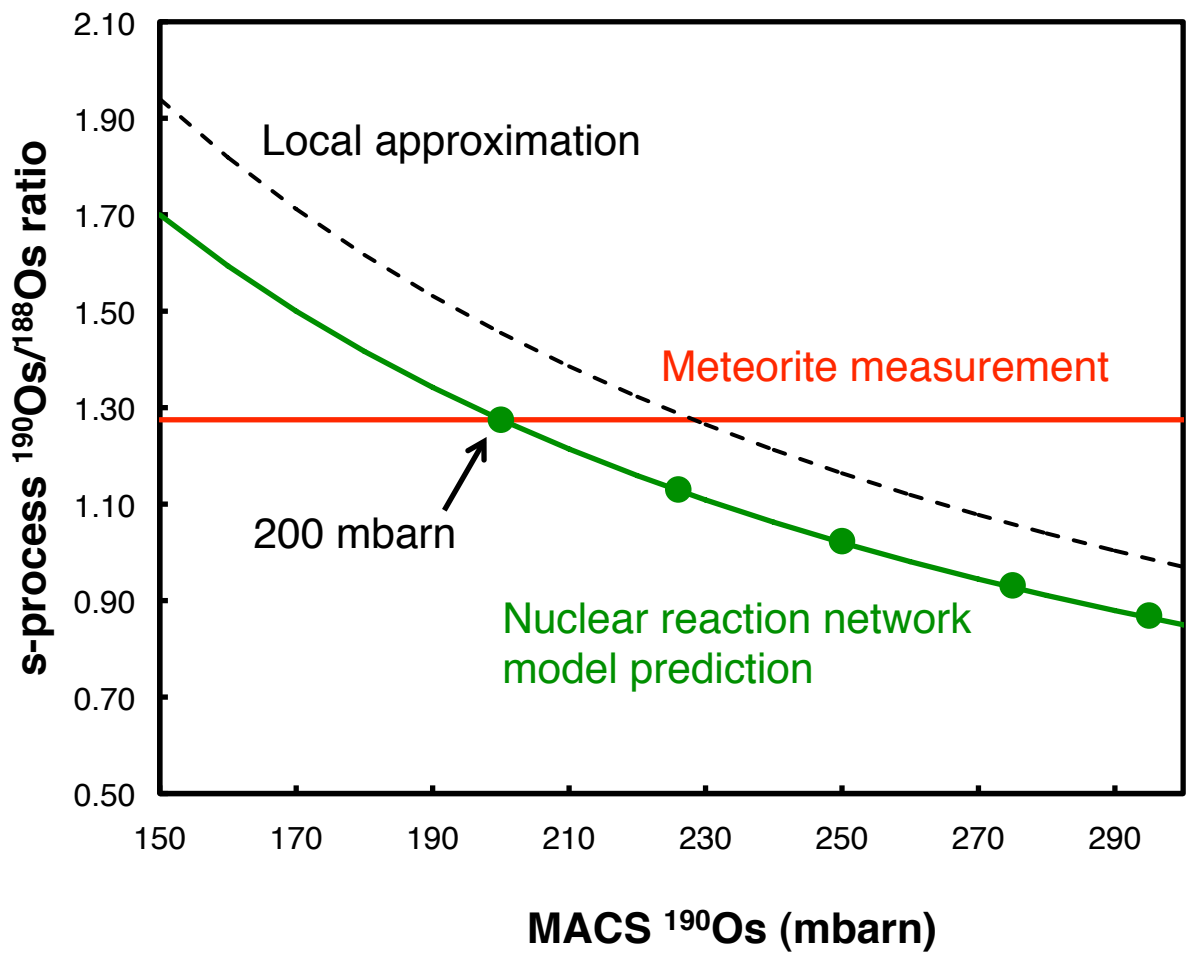

| Leaching Step | Os (ppb) | Pt (ppb) | Re (ppb) | $^{190}$Pt/$^{189}$Os | $^{187}$Re/$^{188}$Os | $^{187}$Os/$^{188}$Os | ε$^{184}$Os | ε$^{186}$Os | ε$^{188}$Os | ε$^{190}$Os |
|---|---|---|---|---|---|---|---|---|---|---|
| 1 | 117 | 93 | 16 | 0.00067 | 0.66 | 0.12479 | 356 ± 124 | -9.57 ± 1.35 | -1.89 ± 0.28 | -1.30 ± 0.17 |
|   |     |    |    |         |      |         | *379 ± 179* | *-10.37 ± 2.11* | *-2.77 ± 0.50* | *-1.19 ± 0.27* |
| 2 | 265 | 243 | 22 | 0.00078 | 0.39 | 0.13056 | -108 ± 72 | 4.84 ± 0.88 | 0.52 ± 0.24 | 0.26 ± 0.14 |
| 3 | 69 | 200 | 2.3 | 0.00244 | 0.16 | 0.12804 | -81 ± 88 | 12.58 ± 0.98 | 1.56 ± 0.23 | 1.05 ± 0.12 |
| 4 | 12 | 21 | 0.5 | 0.00148 | 0.21 | 0.12309 | -58 ± 138 | 6.38 ± 1.30 | 0.93 ± 0.32 | 0.64 ± 0.17 |
|   |    |    |     |         |      |         |           | *7.19 ± 3.61* | *1.75 ± 0.76* | *0.30 ± 0.36* |
| 5 | 20 | 5 | 0.3 | 0.00019 | 0.07 | 0.11922 | 460 ± 146 | -14.08 ± 1.41 | -2.62 ± 0.34 | -1.67 ± 0.18 |
|   |    |   |     |         |      |         |           | *-16.66 ± 9.52* | *-3.86 ± 1.53* | *-1.85 ± 0.80* |
| 6 | 63 | 339 | 6 | 0.00456 | 0.46 | 0.12543 | 44 ± 45 | 1.15 ± 0.53 | 0.39 ± 0.14 | 0.09 ± 0.08 |
| Total (conc.) or weighted average (ratios) | 547 ± 18 | 900 ± 17 | 46 ± 7 | 0.00139 ±0.00005 | 0.41 ±0.06 | 0.12801 ±0.00012 | 34 ± 103 | 1.65 ± 1.44 | 0.01 ± 0.32 | -0.06 ± 0.19 |

| Laboratory | $^{184}$Os/$^{189}$Os | $^{186}$Os/$^{189}$Os | $^{187}$Os/$^{189}$Os | $^{188}$Os/$^{189}$Os | $^{190}$Os/$^{189}$Os |
|---|---|---|---|---|---|
| Durham University | 0.0010650 (85) | 0.0982603 (68) | 0.0932966 (49) | 0.819887 (12) | 1.626405 (11) |
| University of Maryland |  | 0.0982610 (42) | 0.0933004 (57) | 0.819866 (13) | 1.626394 (18) |
| Johnson Space Center | 0.0010734 (48) | 0.0982708 (25) | 0.0933016 (31) | 0.819909 (14) | 1.626411 (23) |

**Table 1** - Upper: Os, Re, and Pt concentrations (blank-corrected) and Os isotopic compositions of the various leachates. Concentrations were calculated by dividing the total quantity of the element released by each leaching step by the mass of the sample. Lines in italics represent preliminary data obtained from small amounts of leachate. Os isotopic data have been corrected for in-run oxygen composition, mass fractionation and potential PtO$_2^-$ and WO$_3^-$ interferences (see SOM for details). Listed uncertainties are in-run 2σ-m precisions. Weighted average uncertainties include effects of individual ratio and concentration uncertainties. Lower: Results of 11 analyses of the University of Maryland standard obtained over the period of measurement compared with data from the same standard obtained at the University of Maryland (Yokoyama et al., 2007) and the Johnson Space Center (online supplemental material of Brandon et al., 2005). Numbers in parentheses indicate 2σ external reproducibility. Fractionation was corrected by normalization to $^{192}$Os/$^{189}$Os = 2.527411. The JSC data have been renormalized to this value (after calculation of $^{184}$Os/$^{189}$Os and $^{187}$Os/$^{189}$Os ratios from ratios listed relative to $^{188}$Os).

| Os isotope | Solar system abundance (4.5 Ga) | p-process | s-process | r-process | cosmoradiogenic |
|---|---|---|---|---|---|
| 184 | 0.000133 | 0.000133 | 0 | 0 | 0 |
| 186 | 0.010727 | 0.00028 | 0.01140 | 0 | $<3 \times 10^{-6}$ |
| 187 | 0.008532 | 0 | 0.00325 | 0 | 0.00528 |
| 188 | 0.089524 | 0 | 0.02498 | 0.06455 | 0 |
| 189 | 0.10915 | 0 | 0.00460 | 0.10455 | 0 |
| 190 | 0.17750 | 0 | 0.02171 | 0.15579 | 0 |
|  |  |  | **0.0318** | **0.1456** |  |
| 192 | 0.27568 | 0 | 0.00315 | 0.27253 | 0 |

**Table 2.** Abundances of Os nuclides, relative to $10^6$ Si atoms. Solar system abundances from Lodders (2003). The p-process abundance of $^{186}$Os is assumed to be ~2.1× the solar system abundance of $^{184}$Os, a p-only isotope (based on solar ratios of p-only isotopes separated by 2 amu, such as $^{156}$Dy and $^{158}$Dy). The MACS values of Mosconi et al. (2006) were used to calculate s-process abundances through $^{188}$Os; heavier Os isotope abundances are based on the MACS values of Bao et al. (2000). Also listed (in bold) is the $^{190}$Os s-process abundance estimated from the $\varepsilon^{190}$Os vs. $\varepsilon^{188}$Os correlation, which corresponds to a $^{190}$Os/$^{188}$Os s-process ratio of 1.275 (see text). This estimate is probably more accurate than the s-process abundance obtained from the measured $^{190}$Os MACS value. The AGB modelling parameters of the FRANEC models used by Zinner et al. (2006) were used for these calculations. The MACS and β-decay rate of $^{185}$W were adjusted within uncertainties to minimize overproduction of s-process $^{186}$Os. Though the calculated s-process abundance exceeds the estimated solar system abundance of this isotope, this discrepancy could perhaps be eliminated by a more complete consideration of all of the uncertainties. Cosmoradiogenic abundance of $^{187}$Os ($^{187}$Os produced by radiogenic decay of $^{187}$Re prior to the formation of the solar system) and r-process abundances of heavier isotopes were obtained by subtraction of s-process from solar system abundances.

## S-1. Analytical techniques

*S-1.1 Leaching procedure*

About 16.5 g of powdered Murchison meteorite were subjected to a sequential leaching procedure at the University of Chicago. Only ultra pure acids and milliQ water were used. The following steps were applied (concentrated acids were used and diluted as indicated):

Step 1)  50 mL acetic acid + 50 mL $H_2O$, 1 day, 20 °C
Step 2)  25 mL $HNO_3$ + 50 mL $H_2O$, 5 days, 20 °C
Step 3)  30 mL HCl + 35 mL $H_2O$, 1 day, 75 °C
Step 4)  30 mL HF + 15 mL HCl + 15 mL $H_2O$, 1 day, 75 °C
Step 5)  10 mL HF + 10 mL HCl, 3 days, 150 °C
Step 6)  2 mL $HNO_3$ + 2 mL HF, 120 °C, 15 hours (applied to only a fraction of Step 5 residue)

Between each leaching step, samples were centrifuged at 5000 g for 30 minutes and rinsed several times in distilled water. A fraction of each leachate was removed for Os analysis; the rest was reserved for isotopic analysis of other elements. The Os fractions of steps 1, 3, 4 and 5 were dried down at low temperature to remove traces of HF or acetic acid. At the Centre de Recherches Pétrographiques et Géochimiques (CRPG-CNRS, Nancy, France), 2 mL HCl were added and the samples were heated at ~ 80 °C for 2 days in closed beakers to redissolve the residues. To insure that no traces of HF remained, steps 4 and 5 were dried down a second time at 70 °C and redissolved in 2 mL HCl. Steps 2 and 6 contained $HNO_3$ and therefore were not dried down. As $HNO_3$ is oxidizing, Os in these fractions would be likely to be in its most oxidized form ($OsO_4$) which is volatile at low temperature and could be lost during drying.

*S-1.2 Chemical separation*

About 10 to 20% of each Os fraction was removed for concentration determinations, while the rest was reserved for high precision Os isotopic measurements. Both aliquots of each leachate were weighed, and the concentration aliquots were spiked with $^{190}$Os, $^{185}$Re, and $^{196}$Pt for isotope dilution analysis. Both the spiked and the unspiked aliquots of steps 1 through 5 were placed in Carius tubes (Shirey and Walker, 1995), and HCl and $HNO_3$ were added to attain total volumes of 3 mL (HCl) and 7 mL ($HNO_3$). The Carius tubes were sealed and heated at 230 °C for 2 days to assure sample digestion and Os oxidation. Complete oxidation is critical to assure spike-sample Os equilibration and to bring Os to its highest oxidation state, which is necessary for the Os extraction chemistry. Step 6 could not be dried down because it contained $HNO_3$, and could not be digested in a Carius tube because it contained HF. Instead, oxidation was effected in teflon digestion vessels by adding ~ 2 g of $CrO_3$ in 2 mL $H_2O$ and 10 mL of concentrated $HNO_3$ to both the spiked and unspiked aliquots, and heating overnight at 90 °C.

Osmium extraction techniques were adapted from those of Birck et al. (1997). After oxidation, the samples were placed with 4 mL $HNO_3$ and 3 mL $Br_2$ in 60 mL teflon digestion vessels, which were sealed with teflon tape to prevent loss of highly volatile $Br_2$. The vessels were placed for 3 hours on a hot plate at ~ 80 °C, allowing scavenging of Os by $Br_2$ as it refluxed through the aqueous phase. The $Br_2$ was then removed, placed in a 15 mL beaker and covered with water to prevent evaporation and impede loss of volatile $OsO_4$. Another 2 mL of $Br_2$ were added to the sample and the refluxing process was repeated to collect any remaining Os. This $Br_2$ was then removed and added to the $Br_2$ from the first reflux. The overlying water was removed and 0.5 mL of HBr were added to reduce the Os to a non-volatile form. The samples were then allowed to dry at room temperature until only the HBr remained. This was dried at ~80 °C, and purified by micro-distillation (Birck, et al., 1997).

The aqueous phase remaining after $Br_2$ removal was dried down and redissolved in 1 M HCl + 10% $Br_2$. Re and Pt were extracted from this solution following method 1 of Rehkämper and Halliday (1997). In brief, the solution was loaded onto anion exchange columns (Bio-Rad AG1X8). Re was then eluted in 8 M $HNO_3$, while Pt was eluted in 13.5 M $HNO_3$. The collected fractions were dried down, redissolved and dried down again in a small amount of concentrated $HNO_3$, then redissolved in $H_2O$ for analysis by ICPMS.

*S-1.3 Mass spectrometry*

Re and Pt concentrations were determined by isotope dilution calculations after analysis by ICPMS (Elan 6000) at the SARM (Service d'Analyse des Roches et des Minéraux) of CRPG. Fractionation was corrected by standard bracketing. Uncertainties (2σ-m precisions) were ~5% for Re, with the exception of fractions 1 and 6, which had uncertainties of ~ 20% due to underspiking. Pt uncertainties varied from 1 to 3%, except for fraction 5 which had an uncertainty of 13% due to its low concentration. Total blanks were 2 pg for Re and 46 pg for Pt. These blank levels do not significantly affect Re and Pt concentrations (except for the Pt concentration of fraction 5).

Os was analyzed as $OsO_3^-$ by negative thermal ionization mass spectrometry (NTIMS) (Creaser, et al., 1991; Volkening, et al., 1991). Concentrations were determined by isotope dilution, based on isotopic compositions measured in peak jumping mode using an ETP electron multiplier on the CRPG Finnigan MAT262 mass spectrometer. Unspiked isotopic compositions were determined by multi-faraday cup measurements on the Durham University Triton mass spectrometer. The procedure used for high precision Os isotopic measurements is described in detail in Luguet et al. (2008). Several points are worth highlighting. The oxygen isotopic compositions used for the oxygen correction were obtained from in-run analyses of masses 239, 240, 241 and 242. Corrections for variations in oxygen composition and for isobaric interferences were performed before mass fractionation corrections. In run fractionation corrections were made assuming a $^{192}Os/^{189}Os$ ratio of 2.5276865, which is coherent with a $^{192}Os/^{188}Os$ ratios of 3.083, the normalization ratio used in recent high precision $^{186}Os$ studies (Brandon, et al., 2006; Walker, et al., 2005). The $^{192}Os/^{189}Os$ ratio, rather than the more commonly employed $^{192}Os/^{188}Os$ ratio, was used for normalization because $^{189}Os$ and $^{192}Os$ display the smallest *s*-process contributions. After analysis, the data were renormalized to a $^{192}Os/^{189}Os$ ratio of 2.527411, using an exponential correction algorithm, to permit comparison with the study of Yokoyama (2007).

Immediately before and after each analysis, 10 blocks of electron multiplier data were obtained for masses 228, 230, 231, 232, and 233. These data were linearly interpolated and used to correct for potential interferences of $PtO_2^-$ and $WO_3^-$ on $^{184}OsO_3^-$ and $^{186}OsO_3^-$, and of $ReO_3^-$ on $^{187}OsO_3^-$. For $^{184}OsO_3^-$, which is a very small peak, these interferences were quite large, ranging from 1 to 10% of the mass 232 intensity, mostly due to the contribution of $^{198}Pt^{18}O^{16}O^-$ (Table S-1). Nevertheless, given the very wide range of variation of the data (about 600 epsilon units), the results are probably meaningful. Both the good agreement between the preliminary and final analyses of the step 1 leachate, despite differing interference corrections (8 and 4%, respectively), and the marked correlation between $\varepsilon^{184}Os$ and $\varepsilon^{188}Os$, argue for the reliability of the corrected data. The only sample that plots significantly off this correlation is the step 2 leachate, which had the largest interference correction. The use of a linear interpolation function to estimate the isobaric interferences at the time of each Os measurment may have led to a slight overestimation of the correction, since the interference peaks, like the observed Os signal, probably in fact decayed almost exponentially rather than linearly. This effect may have been significant for step 2, for which the Os signal displayed the most marked exponential decay, but was probably much smaller or insignificant for the other steps. This may explain why step 2 plots to the low $^{184}Os$ side of the correlation.

For $^{186}OsO_3^-$ (mass 234), the potential interferences are from $^{198}Pt^{18}O_2^-$ and $^{186}WO_3^-$. The $PtO_2^-$ interference was usually < 0.01 epsilon units, and always < 0.1 epsilon, and thus was smaller than the statistical uncertainty of the sample measurements. The $^{186}WO_3^-$ interferences, calculated from $^{183}WO_3^-$ (mass 231 after subtracting the $^{198}Pt^{17}O^{16}O^-$ contribution), were more significant. For the final analyses, the corrections for $^{186}WO_3^-$ on $^{186}Os^-$ ranged between 0.2 and 1.0 epsilon units. The $^{186}WO_3^-$ corrections for the preliminary analyses (indicated in italics in Table 1) were much higher (2 epsilon units for step 1, 5 for step 4, and 43 for step 5). The fact that the new results for step 5, as well as those for steps 1 and 4, agree within error with the original results, despite the very large $^{186}WO_3^-$ correction applied to the preliminary analysis, indicates that the $^{186}WO_3^-$ correction was performed properly. The source of the very high $WO_3^-$ interferences in the preliminary analyses may have been the Carius tubes, as these are made of borosilicate glass which is rich in W. The sample (step 5 preliminary data) with the very large $WO_3^-$ correction was stored in a Carius tube for many months before analysis, while the samples with small W corrections (< 0.5 epsilon units) were analyzed soon after digestion (new analyses of steps 1, 4 and 5). Step 6, which was not put in a Carius tube at all, also had only a small $WO_3^-$ correction. Significant $WO_3^-$ interferences have never been observed in samples digested in quartz vessels in a high pressure asher at Durham University.

Starting beam intensities for $^{186}OsO_3^-$ ranged between 18 and 93 mV, for initial Os quantities (before separation chemistry) ranging from 29 to 80 ng. These are lower than the >80 mV typically obtained for 30 ng loads of Os standard on the Durham Triton, and most probably reflect incomplete chemical yields. The low beam intensities explain the relatively high uncertainties and interference corrections of some of the measurements. Total Os blanks, from the Carius tube step on, were about 0.5 pg, and thus are insignificant relative to the sample sizes. It is difficult to characterize the Os blanks of each leaching step, but given that purified reagents were used, they were most probably insignificant relative to the large quantities of Os in the samples. Since the blank composition must be terrestrial, any blank contribution would decrease the magnitude of the observed anomalies. The blank will have no effect on the slopes of the correlations between various isotopes as these reflect mixing lines with terrestrial Os.

The Os standard solution used in Durham was obtained from the University of Maryland (standard UMd) and is the same as that used by both Yokoyama et al. (2007) and Brandon et al. (2005). Results from the three laboratories (Table 1 of main text) are generally in good agreement, and the standard reproducibility (2σ) is similar. Using the more conventional $^{192}Os/^{188}Os$ normalization, the Durham results correspond to $^{186}Os/^{188}Os$ and $^{187}Os/^{188}Os$ ratios of 0.1198427 (67) and 0.113787 (8) respectively (Luguet, et al., 2008) (2σ uncertainties in parentheses). Os isotopic compositions of the samples are presented in epsilon units, where:

$$\varepsilon^i Os = 10^4 * [(^iOs/^{189}Os)/(^iOs/^{189}Os_{std}) - 1] \text{ and } i = 186, 187, 188, \text{ or } 190$$

The UMd standard isotopic ratios determined in Durham (Table 1 of main text) were used as the reference values for all isotopes except $^{186}Os$. Since this standard has a radiogenic $^{186}Os/^{189}Os$ composition, we used instead the average chondritic $^{186}Os/^{189}Os$ ratio (0.0982524) determined by Yokoyama et al. (2007) to calculate $\varepsilon^{186}Os$, which facilitated comparison with previous studies. Use of the UMd ratio would result in $\varepsilon^{186}Os$ values about 0.6 units lower than those listed.

| Leachate | PtO$_2$ interferences | WO$_3$ interferences | Total estimated interference (%) |
|---|---|---|---|
| 1 (preliminary) | 5.4 | 2.2 | 7.7 |
| 1 | 3.6 | 0.6 | 4.1 |
| 2 | 7.9 | 1.9 | 9.8 |
| 3 | 4.9 | 1.2 | 6.1 |
| 4 | 6.0 | 0.8 | 6.8 |
| 5 | 2.2 | 0.5 | 2.6 |
| 6 | 0.9 | 0.3 | 1.2 |

**Table S-1.** PtO$_2$ and WO$_3$ interferences (in % relative to the mass 232 peak) estimated by linear interpolation of electron multiplier data for each leachate step.

*S-1.4 Factors complicating interlaboratory comparisons*

Comparison with the previous high precision Os isotopic results of Brandon et al. (2005) and Yokoyama et al. (2007) is complicated by small differences in analytical procedure and normalization protocols. 1) Here and in the Yokoyama et al. study, fractionation was corrected assuming a $^{192}Os/^{189}Os$ ratio of 2.527411, while Brandon et al. assumed a ratio of 2.527685. Fortunately, use of the epsilon notation essentially removes the effects of this different normalization. 2) In our study, oxygen corrections were made using in-run oxygen composition measurements. In the Yokoyama et al. study, the oxygen isotopic composition determined by Nier (1937) was assumed, while that used in the Brandon et al. study was not specified. 3) In our study, potential interferences from PtO$_2$ and WO$_3$ were monitored and corrected based on SEM measurements before and after each run, which were interpolated to estimate the interference at the time of each Os measurement.  In the other studies, potential interferences were monitored by scanning the relevant masses before or after the runs.  This procedural difference may be particularly important for $^{184}$Os measurements.  4) In our study, the data are normalized relative to our measured Os isotopic ratios for the University of Maryland standard (except for $^{186}$Os).  In the other studies, the data are normalized relative to the average bulk compositions obtained for a group of 5 H-chondrites (Brandon et al.) or of 9 chondrites of various classes (Yokoyama et al.).  5) In the other studies, the discussion is based on initial $\varepsilon^{186}$Os values, which are obtained by correcting the measured $^{186}$Os/$^{189}$Os ratios for 4.56 Ga of $^{190}$Pt decay based on the measured Pt/Os ratios, and comparing these with the average chondritic $^{186}$Os/$^{189}$Os ratio at 4.56 Ga defined by their measurements.  Such correction is impossible for our samples since, as discussed in the main text, the Pt/Os ratios of the leachates have probably been modified by the leaching procedure itself, and are thus inappropriate to use for the radiogenic correction.  In any case, as shown by Brandon et al. (their supplemental material), $\varepsilon^{186}$Os values obtained by comparison with the present-day chondritic average are nearly identical to those obtained by correcting the sample $^{186}$Os/$^{189}$Os ratios to 4.56 Ga and comparing with the average chondritic $^{186}$Os/$^{189}$Os at 4.56 Ga.  It is thus possible to compare age corrected and uncorrected $\varepsilon^{186}$Os values. 6)  In our study, as well as in the Yokoyama et al. and Brandon et al. studies, the fractionation correction is performed after the oxygen correction using an exponential law, so in this respect the results are directly comparable. Nevertheless, this procedure implicitly assumes that all of the fractionation occurs before or during formation of the OsO$_3^-$ ions, which may not be the case.  If fractionation actually occurs after molecular ion formation, it would be more appropriate to perform the fractionation correction before the oxygen correction.  This would lead to minor changes in the calculated $\varepsilon^i$Os values.  For example, for a fractionation factor ($\alpha$ value) of 0.5, the $\varepsilon^{186}$Os value would be about 0.35 epsilon units higher if calculated before the oxygen correction.

In the present case, the observed anomalies are so large that discrepancies caused by interlaboratory differences in data acquisition and correction procedures are insignificant. However, in order to look at more subtle variations, it is essential to develop an internationally accepted standard and a uniform data reduction scheme to facilitate interlaboratory comparison of high precision Os isotopic measurements.

## S-2. Derivation of equation for *s*-process $^{190}$Os/$^{188}$Os

From eq. (1) in the main text, for $^{190}$Os,

$$\varepsilon_{Os}^{190} = \frac{\rho_{Os}^{190} - \rho_{Os}^{192}\mu_{Os}^{190}}{\rho_{Os}^{188} - \rho_{Os}^{192}\mu_{Os}^{188}} \varepsilon^{188}Os \tag{A1}$$

where $\rho_{Os}^i = \dfrac{{}^i Os/{}^{189}Os_s}{{}^i Os/{}^{189}Os_{ss}} - 1$ and $\mu_{Os}^i = \dfrac{i-189}{192-189}$, from equations (2) and (3) respectively.

Let $m$ be the slope of the correlation of $\varepsilon^{190}$ vs. $\varepsilon^{188}$, i.e., $m = \varepsilon^{190}/\varepsilon^{188}$. Using the expression for $\rho^i$, $m$ becomes:

$$m = \frac{\dfrac{(^{190}\text{Os}/^{189}\text{Os})s}{(^{190}\text{Os}/^{189}\text{Os})ss} - 1 - \mu^{190}\left(\dfrac{(^{192}\text{Os}/^{189}\text{Os})s}{(^{192}\text{Os}/^{189}\text{Os})ss} - 1\right)}{\dfrac{(^{188}\text{Os}/^{189}\text{Os})s}{(^{188}\text{Os}/^{189}\text{Os})ss} - 1 - \mu^{188}\left(\dfrac{(^{192}\text{Os}/^{189}\text{Os})s}{(^{192}\text{Os}/^{189}\text{Os})ss} - 1\right)} \qquad (A2)$$

Multiplying numerator and denominator by $\dfrac{(^{189}\text{Os}/^{188}\text{Os})_s}{(^{189}\text{Os}/^{188}\text{Os})_{ss}}$ we obtain:

$$m = \frac{\dfrac{(^{190}\text{Os}/^{188}\text{Os})s}{(^{190}\text{Os}/^{188}\text{Os})ss} - \dfrac{(^{189}\text{Os}/^{188}\text{Os})s}{(^{189}\text{Os}/^{188}\text{Os})ss} - \mu^{190}\left[\dfrac{(^{192}\text{Os}/^{188}\text{Os})s}{(^{192}\text{Os}/^{188}\text{Os})ss} - \dfrac{(^{189}\text{Os}/^{188}\text{Os})s}{(^{189}\text{Os}/^{188}\text{Os})ss}\right]}{1 - \dfrac{(^{189}\text{Os}/^{188}\text{Os})s}{(^{189}\text{Os}/^{188}\text{Os})ss} - \mu^{188}\left[\dfrac{(^{192}\text{Os}/^{188}\text{Os})s}{(^{192}\text{Os}/^{188}\text{Os})ss} - \dfrac{(^{189}\text{Os}/^{188}\text{Os})s}{(^{189}\text{Os}/^{188}\text{Os})ss}\right]} \qquad (A3)$$

or,

$$m = \frac{\dfrac{(^{190}\text{Os}/^{188}\text{Os})s}{(^{190}\text{Os}/^{188}\text{Os})ss} - \dfrac{(^{189}\text{Os}/^{188}\text{Os})s}{(^{189}\text{Os}/^{188}\text{Os})ss}(1-\mu^{190}) - \mu^{190}\left(\dfrac{(^{192}\text{Os}/^{188}\text{Os})s}{(^{192}\text{Os}/^{188}\text{Os})ss}\right)}{1 - \dfrac{(^{189}\text{Os}/^{188}\text{Os})s}{(^{189}\text{Os}/^{188}\text{Os})ss}(1-\mu^{188}) - \mu^{188}\left(\dfrac{(^{192}\text{Os}/^{188}\text{Os})s}{(^{192}\text{Os}/^{188}\text{Os})ss}\right)} \qquad (A4)$$

let $R^i_{\text{Os}} = \dfrac{(^i\text{Os}/^{188}\text{Os})s}{(^i\text{Os}/^{188}\text{Os})ss}$. After rearranging terms,

$$\frac{(^{190}\text{Os}/^{188}\text{Os})s}{(^{190}\text{Os}/^{188}\text{Os})ss} = m\left[1 - R^{189}_{\text{Os}}(1-\mu^{188}) - \mu^{188}R^{192}_{\text{Os}}\right] + R^{189}_{\text{Os}}(1-\mu^{190}) + \mu^{190}R^{192}_{\text{Os}}$$

or,

$$\frac{(^{190}\text{Os}/^{188}\text{Os})s}{(^{190}\text{Os}/^{188}\text{Os})ss} = m + R^{189}_{\text{Os}}(1-\mu^{190} - m + m\mu^{188}) + R^{192}_{\text{Os}}(\mu^{190} - m\mu^{188}) \qquad (A5)$$

The slope of the correlation line, $m$, in Fig. 8 is 0.651. From eq. 3, $\mu^{190} = 1/3$, while $\mu^{188} = -1/3$. Using the data in Table 2, $R^{189} = 0.1525$ and $R^{192} = 0.0408$. Thus,

$$\frac{(^{190}\text{Os}/^{188}\text{Os})s}{(^{190}\text{Os}/^{188}\text{Os})ss} = 0.651 - 0.0307 + 0.0225 = 0.643.$$

Using $^{190}\text{Os}/^{188}\text{Os})ss = 1.983$ (Lodders, 2003), the $s$-process $^{190}\text{Os}/^{188}\text{Os}$ ratio is estimated to be 1.275 ±0.041 (95 % confidence interval)

**References**

Birck, J.-L., Roy Barman, M. and Capmas, F., 1997. Re-Os isotopic measurements at the femtomole level in natural samples, Geostandards Newsletter 21, 19-27.

Brandon, A.D., Humayun, M., Puchtel, I.S., Leya, I. and Zolensky, M., 2005. Osmium isotope evidence for an s-process carrier in primitive chondrites, Science 309, 1233-1236.

Brandon, A.D., Walker, R.J. and Puchtel, I.S., 2006. Platinum-osmium isotope evolution of the Earth's mantle: Constraints from chondrites and Os-rich alloys, Geochim. Cosmochim. Acta 70, 2093-2103.

Creaser, R.A., Papanastassiou, D. and Wasserburg, G.J., 1991. Negative thermal ion mass spectrometry of osmium, rhenium, and iridium, Geochim. Cosmochim. Acta 55, 397-401.

Lodders, K., 2003. Solar system abundances and condensation temperatures of the elements, Astrophys. J. 591, 1220-1247.

Luguet, A., Nowell, G.M. and Pearson, D.G., 2008. $^{184}Os/^{188}Os$ and $^{186}Os/^{188}Os$ measurements by negative thermal ionisation mass spectrometry (NTIMS): Effects of interfering element and mass fractionation corrections on data accuracy and precision, Chem. Geol. 248 (special issue HSE Geochemistry), 342-362.

Rehkämper, M. and Halliday, A.N., 1997. Development and application of new ion-exchange techniques for the separation of the platinum group and other siderophile elements from geological samples, Talanta 44, 663-672.

Shirey, S.B. and Walker, R.J., 1995. Carius tube digestion for low-blank rhenium-osmium analysis, Anal. Chem. 67, 2136-2141.

Volkening, J., Walczyk, T. and Heumann, K.G., 1991. Osmium isotope determinations by negative thermal ionization mass spectrometry, Int. J. Mass Spec. Ion Phys. 105, 147-159.

Walker, R.J., Brandon, A.D., Bird, J.M., Piccoli, P.M., McDonough, W.F. and Ash, R.D., 2005. $^{187}Os$-$^{186}Os$ systematics of Os-Ir-Ru alloy grains from southwestern Oregon, Earth Planet. Sci. Lett. 230, 211-226.

Yokoyama, T., Rai, V.K., Alexander, C.M.O.D., Lewis, R.S., Carlson, R.W., Shirey, S.B., Thiemens, M.H. and Walker, R.J., 2007. Osmium isotope evidence for uniform distribution of s- and r-process components in the early solar system, Earth Planet. Sci. Lett. 259, 567-580.